\documentclass[preprint,pra,aps,showpacs,preprintnumbers,amsmath,amssymb,eqsecnum]{revtex4}

\usepackage{amsmath}
\usepackage{amsfonts}

\usepackage{graphicx}

\def\ra{{\rangle}}

\def\e{{\epsilon}}
\def\eps{{\epsilon}}

\def\ih{{ \frac{i}{\hbar} }}

\def\half{\frac{1}{2}}
\def\pp{\prime\prime}

\pacs{03.65.Yz, 03.65.Nk, 03.65.Xp}

\newcommand\beq{\begin{equation}}
\newcommand\eeq{\end{equation}}
\newcommand\bea{\begin{eqnarray}}
\newcommand\eea {\end{eqnarray}}

\begin{document}

\title{Suppression of quantum-mechanical reflection by environmental decoherence}
\author{D. J. Bedingham}
\author{J. J. Halliwell}
\affiliation{Blackett Laboratory \\ Imperial College \\ London SW7 2BZ \\ UK}

\begin{abstract}
In quantum mechanics an incoming particle wave packet with sufficient energy will
undergo both transmission and reflection when encountering a barrier of lower energy,
but in classical mechanics there is no reflection, only transmission. In this paper
we seek to explain the disappearance of quantum-mechanical reflection in the quasi-classical
limit, using the standard machinery of decoherence through environmental interaction.
We consider two models. In the first, the incoming particle is classicalized by coupling
to an environment and modelled using a standard master equation of Lindblad form
with Lindblad operator proportional to position (the simplest version of
quantum Brownian motion). We find, however, that suppression of reflection is achieved
only for environmental interaction so strong that large fluctuations in momentum are generated
which blurs the distinction between incoming and reflected wave packets. This negative conclusion also holds for a complex potential which has similar implications
for attempts to understand the suppression of the Zeno effect using the same mechanism
(discussed in more detail in another paper).
A different master equation with Lindblad operator proportional to momentum is shown
to be successful in suppressing reflection without large fluctuations but such a
master equation is unphysical. We consider a second model in which the barrier is
modeled quantum-mechanically by a massive target particle coupled to an environment
to maintain it in a quasi-classical state. This avoids the fluctuations problem since
the incoming particle is not coupled to the environment directly. We find that reflection
is significantly suppressed as long as the decoherence timescale of the target particle
is much smaller than certain characteristic scattering timescales of the incoming particle, or equivalently, as long as the velocity fluctuations in the target are larger than the velocity of the incoming particle.
\end{abstract}

\maketitle

\section{Introduction}

\subsection{Opening Remarks}

There has been considerable interest over the years in questions relating to the
emergence of classical behaviour from quantum theory. Studies in this area focus
on a number of related issues, in particular, suppression of interference and entanglement through interaction
with an environment, finding the conditions under which probability sum rules for
histories are satisfied, and deriving the effective classical equations of motion
from an underlying quantum system
\cite{Har6,JoZ,Hal8,Hal00,Zur1,Zur2,Zur3,BrHa,
GH2,GH1,Gri,Omn,Hal1,DoH,HaDo}.

However, despite the considerable amount of work in this area there still remain a number of
simple quantum effects whose classical limit is not yet well-understood. A particular
example is quantum-mechanical reflection from a simple barrier.
The aim of this paper is to explore the classical limit of this situation in detail.
We do not have a particular experimental situation in mind, but rather, we are interested
in understanding the general mechanism whereby this classical limit comes about.

\subsection{The Problem. Model I}

We consider a particle in one dimension in an incoming wave packet state of positive momentum
with average energy  $E$ approaching a barrier of height $V_0$. Classically, if
$ E > V_0$ the particle will continue over the barrier but in quantum
mechanics there is both transmission and reflection. How do we understand the
disappearance of quantum-mechanical reflection in the classical limit?

Consider first the naive classical limit $ \hbar \rightarrow 0 $ in the scattering amplitude.
It is known that the reflection amplitudes do not go to zero in this limit for simple step
function potentials and some smoothing of the edges is required \cite{LaLi}. For example,
suppose that the potential consists of a smeared window function,
\beq
V(x) =  V_0 \int_{-\infty}^{\infty} d \bar x \ \frac{ 1} { (2 \pi a^2)^{1/2} } \exp
\left(-\frac {(x- \bar x)^2}{2a^2}\right)
f_L (\bar x),
\label{pot}
\eeq
where $f_L(\bar x)$ is a window function on the range $[-L,L]$.
Then for $ E > V_0$, the  Born approximation gives for the reflected wavefunction amplitude squared
\beq
|\psi^{\rm ref}_{\infty}(p)|^2  = \frac{2\pi  V_0^2 m^2}{\hbar p^2}
\exp\left(-\frac{4a^2 p^2}{\hbar^2}\right)\ f_L^2(2p) |\psi_{-\infty}(-p)|^2,
\label{Uref}
\eeq
where $f_L(p)$ denotes the Fourier transform of $f_L(x)$
This is indeed small for $ pa \gg \hbar $ and in this sense goes to zero as $\hbar \rightarrow 0 $.
However, this naive limit is clearly insufficient for a comprehensive demonstration of emergent classicality.

The general wisdom in the area of emergent classicality is that coupling a quantum
system to a thermal environment produces decoherence which tends to suppress all quantum
effects \cite{JoZ,Zur1,Zur2,Zur3}. This general approach has been shown to work well in many different situations
so we will explore it here. We therefore consider a more complicated version of the
above system in which the incoming particle is also coupled to a thermal environment as it
undergoes scattering off a potential barrier. We make the realistic assumption
that the effect of the environment is Markovian, so the density operator will evolve
according to a Lindblad master equation \cite{Lin} of the form
\beq
\dot \rho = - \ih [ H_0 + V, \rho] - \frac {1}{2} [ L, [L, \rho]],
\label{lin}
\eeq
where $H_0$ is the free Hamiltonian and the potential $V(x)$ is given by Eq.(\ref{pot}).
The physically interesting case of quantum Brownian motion with negligible
dissipation is obtained with the choice $L = (2D / \hbar^2)^{1/2} \ \! \hat x$.
However it will also be of interest to consider other choices for $L$.

It is also useful to consider this dynamics in the Wigner picture \cite{Wig,KiNo} in which the Wigner
function $W(p,x)$ evolves according to
\bea
\frac {\partial W} {\partial t} = &-& \frac {p}{m} \frac {\partial W} {\partial x}
+ V'(x) \frac {\partial W} {\partial p}
+ D \frac {\partial^2 W} {\partial p^2}
\nonumber \\
&+& \sum_{k=1}^{\infty} \left( \frac {i \hbar }{ 2 } \right)^{2k} { 1
\over (2k+1)!} V^{(2k+1)}(q) \ \frac { \partial^{2k+1} W }{ \partial p^{2k+1} }.
\label{Wig1}
\eea
The traditional understanding of this system is that the diffusion produced
by the environment spreads out the Wigner function and as a result the higher
order quantum terms involving powers of $\hbar^2$ are strongly suppressed. The Wigner function
thus evolves according to a classical stochastic theory of a particle in a potential.
This classical stochastic theory can be an approximately deterministic theory
if the fluctuations produced by the diffusive term are small enough, since an initial state strongly peaked about a point in phase space will remain strongly peaked.
In earlier
applications, this has been found to be possible if the particle is sufficiently massive, but this point needs particular attention in our case.
Furthermore, the Wigner function typically becomes positive after a short time \cite{DiKi} so
may therefore be loosely interpreted as a phase space distribution function.
This is the general way in which quantum systems become quasi-classical through decoherence.

For the reflection process studied here, the classical stochastic theory described
by Eq.(\ref{Wig1}) with quantum terms omitted has no reflection, so our goal is achieved
if the stochastic effects are negligible. However, what we find in this model
is that for the physically relevant master equation with $L \propto \hat x$, reflection can be significantly
suppressed by taking $D$ to be large enough,
but this generates fluctuations in the momentum which are so large that they blur the distinction between
positive and negative momenta. This means
that even though there is no actual reflection, there is still diffusion
into negative momenta, which could be as significant as the reflection effect we
are trying to suppress. Differently put, for the environment to be strong enough to significantly suppress
reflection, the integrity of the incoming wave packet is destroyed.

The underlying reason for this difficulty is that reflection is in essence an interference effect in momentum space. The Lindblad equation with $L \propto \hat x$ is well-known to produce
very rapid decoherence in position space, but this takes a longer time, through
the action of the Hamiltonian, to rotate into momentum space decoherence. The momentum
space decoherence time can be made short by making $D$ sufficiently large, but this then produces
unacceptably large fluctuations.
This explanation of the difficulty is confirmed by using instead a Lindblad master
equation with $L \propto \hat p$. We find in this case that reflection is easily suppressed without incurring
large fluctuations. However, we know of no physical situation that produces such
an equation under realistic conditions. We conclude that quantum-mechanical reflection
is not obviously suppressed by the obvious (and usually very effective) choice of decoherence
mechanism.

The possible suppression of reflection through interaction with an environment has been considered
previously and there are some experimental indications that scattering intensity is reduced by the
presence of a thermal environment \cite{ChTe}. The environment was modeled using a Lindblad
master equation with $L \propto \hat p $ in Ref.\cite{ChTe} and it was noted that this reduces reflection but
a detailed calculation as not carried out. Also, the difficulties noted here with the physically realistic
case $L \propto \hat x $ were not discussed.

\subsection{Relation to the Zeno Effect}

The situation described above is closely related to the
the Zeno effect \cite{Zeno,Sch2}, and in particular its possible suppression through decoherence, described in another paper \cite{BeHa2} and we briefly note
the connections here. Again we consider an incoming wave packet $ | \psi \rangle
$
concentrated around positive momentum,
in one dimension, and we suppose that the system is measured at time intervals
$\eps$ to see if it lies in the negative $x$-axis. The state at time $\tau = n \eps$
after this sequence of unitary evolutions interspersed with projections $P =\theta
( - \hat x) $ is given
by
\beq
|\psi_\tau \rangle = \left( P \exp \left( - \ih H \eps  \right) \right)^n|\psi \rangle.
\label{psitau}
\eeq
As is well-known, in the limit $n \rightarrow \infty$, $\eps \rightarrow 0 $ with
$\tau$ fixed, the evolution becomes unitary in the subspace of states with support
in $x<0$,
\beq
| \psi_\tau \rangle = P \exp \left( - \ih PHP \right) | \psi \rangle.
\eeq
This means that the incoming wave packet is totally reflected. For small but
finite $\eps$, there is some reflection, with the rest of the state absorbed.
An interesting question is then the extent to which this reflection is suppressed
through decoherence, leaving only absorbtion (the classical limit of the above process).

A useful way to analyze this question, which relates to the model of the previous
subsection, is to make use of the recent results of Refs.\cite{Ech,HaYe3}, where is was shown that
the string of projectors in Eq.(\ref{psitau}) is well-approximated by evolution in the presence
of a complex potential
\beq
| \psi_\tau \rangle \approx P \exp \left( - \ih H \tau - \frac{V_0} {\hbar} \theta (\hat x ) \right)
| \psi \rangle,
\eeq
where $V_0$ is of order $ \hbar / \eps $. Complex potentials of this type appear in many places \cite{complex} and this sort of construction also arises in connection with the arrival time problem \cite{time,All,HaYe1,HaYe2}.

The scattering behaviour of incoming wave packets off a complex potential of this
form is very similar to the real potential case discussed above. The analysis outlined
in the previous section may therefore also be used to investigate the suppression
of the Zeno effect. However, we quickly draw the same negative conclusion as the
previous section: the Zeno effect cannot be significantly suppressed by decoherence if the master equation
has $L \propto \hat x$, since the fluctuations will be too large. Again the environment produces such large fluctuations that the final state includes significant negative momentum and we do not obtain the expected classical limit of total absorbtion of the incoming state.

\subsection{Quantum State Diffusion Approach}

In the following sections we will analyze the above models using the density matrix evolution equation Eq.(\ref{lin}). However, an alternative and very useful way of analyzing this system is to use the quantum state diffusion picture
\cite{GP1,GP2,GP3,Dio8,Dio2,Dio3,DGHP,HaZo1,HaZo2}.
In this picture,
the density operator
$\rho$ satisfying Eq.(\ref{lin}) is regarded
as a mean over a distribution of pure state
density operators,
\beq
\rho = M | \psi \rangle \langle \psi |,
\label{rhoM}
\eeq
where $M$ denotes the mean,
with the pure states evolving according to
the non-linear stochastic Langevin-It\^o equation,
\bea
d |  \psi \rangle &=& -\ih (H_0 +V) |\psi \rangle dt  - \half  \left(
L - \langle L \rangle \right)^2 | \psi \rangle\ dt
\nonumber \\
&+& \left( L - \langle L \rangle \right) | \psi \rangle \ d B_t,
\label{qsd1}
\eea
for the normalized state vector $| \psi \ra $. (The equation is given here for a real potential and readily generalizes to the case of a complex potential).
Here, the $d B_t$ are independent real differential random
variables representing a real Wiener process and satisfying
\bea
dB_t^2 &=& dt,
\\
dB_t dB_{t'} &=& 0, \ {\rm if} \ t \ne t'.
\eea

Decoherence processes in the language of QSD have been analyzed in detail in previous works \cite{HaZo1,HaZo2}. For our purposes, the salient features are as follows. In the case $L = (2D / \hbar^2)^{1/2} \hat x$
and $V=0$, it has been shown \cite{Dio3} that an arbitrary initial state localizes to a Gaussian $ |\psi_{pq} \rangle $
of the form
\beq
\psi_{pq} (x) = \frac {1} { ( 2 \pi\sigma_q^2 )^{1/4} }\exp \left( - (1-i) \frac{ (x-q)^2 }{ 4 \sigma_q^2 } + \ih p x \right),
\label{loc}
\eeq
where the widths are given by
\beq
\sigma^2_q = \left( \frac {\hbar^3} {8 m D} \right)^{1/2}, \ \ \ \
\sigma_p^2 = ( 2 m \hbar D)^{1/2}.
\label{widths}
\eeq
This localization takes place on the localization timescale
\beq
t_{loc} = \left( \frac{m \hbar} {D} \right)^{1/2},
\label{tloc}
\eeq
which is also known to be the timescale on which the Wigner function becomes positive in the evolution described by Eq.(\ref{Wig1}) \cite{DiKi}. Through Eq.(\ref{rhoM}), this means that the corresponding density matrix tends, on the localization timescale, to a mixed state of the approximately diagonal form
\beq
\rho = \int dp dq f(p,q,t) | \psi_{pq} \rangle \langle \psi_{pq} |,
\label{rholoc}
\eeq
where $f(p,q,t)$ is a positive phase space distribution function which has the general form of a smeared Wigner function \cite{HaZo1}. Hence the picture we have is that an arbitrary initial state ends up in a stochastic ensemble of Gaussian states each one localized about a classical stochastic path.
An analogous story also holds for the case $L \propto \hat p$ but in that case $\rho$ tends to a form that is diagonal in momentum.

The QSD picture is another convenient way of establishing the properties of Model I outlined above. The key point is that in the presence of a reflecting potential, the localization time has to be sufficiently short in comparison to the timescale on which reflected wave packets start to form but, as we shall show, this can only be achieved at the expense of unacceptably large fluctuations.

\subsection{A Modified Problem. Model II}

The above models suggest that interaction of the
incoming particle with a decohering environment is not in fact the mechanism whereby
reflection off a barrier is suppressed and we therefore need to look elsewhere
for the mechanism. We therefore focus instead on the barrier, rather than the particle
and note that a barrier is really an idealized model for another system which will be subject to the laws of quantum theory. Our conjecture is that, loosely
speaking, classicalizing this quantum model of a barrier will lead to the suppression
of reflection.

We thus consider a second model in which the incoming
particle interacts through a potential $V$ with a much more massive particle
initially positioned at the origin.
These two particles are described by the
Hamiltonian
\beq
H = \frac {p^2} {2m} + \frac {P^2} {2M} + V (x-X),
\label{ham}
\eeq
where $x,p$ denote the light particle phase space coordinates and those of the massive particle are denoted by $X,P$.
Clearly when $X$ and $P$ are set to zero we recover the usual elementary model of scattering
off a potential $V(x)$ whose classical limit we seek to understand. Instead of coupling
the environment to the light incoming particle, we couple it to the massive particle. We suppose that the massive particle alone is described by a Lindblad master
equation with $L$ proportional to position, the physically realistic case. The master
equation for the density operator of the combined two particle system is therefore taken to be
\beq
\dot \rho = - \ih [ H, \rho] - \frac {D} {\hbar^2}
[ \hat X, [\hat X, \rho]],
\label{lin2}
\eeq
where $H$ is given by Eq.(\ref{ham}). We take the initial state to be
$\rho_0 = | \psi \rangle \langle \psi | \otimes \rho_M $, where the massive particle initial
state $\rho_M$ is a Gaussian concentrated around $X=0=P$.

The key point of this model is that, since the environment does not interact directly
with the light particle, its interaction can be made very strong without giving the
light particle large fluctuations. We thus avoid the fluctuations problem of our
first model. We will indeed find in this
model that quantum-mechanical reflection is suppressed without large fluctuations of the light particle, under suitable conditions.

\subsection{How Reflection is Suppressed}

We briefly outline the specific mechanism whereby reflection is suppressed in Models I and II. As indicated, in both models, it is at a heuristic level because, in the Wigner evolution equation Eq.(\ref{Wig1}), the effect of the environment is to spread out the Wigner function in phase space and this suppresses the non-classical terms (involving the higher derivatives), leaving a classical stochastic theory in which there is no reflection. Hence it is the spreading effect of the environment
that is important.

More precisely, what we find in detailed calculations is as follows. In both models, the effect of the environment is to produce a spread in the reflected momentum of the light particle so that the usual $ \delta$-function $ \delta (p + \bar p)$ in the amplitude, indicating exact reflection, becomes very broad. This spreading alone is not sufficient to suppress reflection. What is also important is that some of the reflected momentum is spread into the positive momentum regime so is not reflected any more, and the rest of it is spread into very large negative values where it is suppressed by terms of the form appearing in the scattering amplitude Eq.(\ref{Uref}), mainly by the term coming from the smearing of the potential but also by prefactors in the scattering amplitude. (This is the sense in which
$ p a \gg \hbar $ is achieved in Eq.(\ref{Uref}) but without having to take $\hbar
\rightarrow 0 $).
As a result the total probability of any final negative momenta is very small, for sufficiently strong environment.
As indicated, suppression occurs in both Models I and II, but only in Model II are the associated fluctuations acceptably small.

\subsection{Outline of the Rest of this Paper}

The above subsections have sketched mainly in simple physical terms the overall picture -- coupling a scattering particle directly to an environment, Model I, does not seem to be effective at suppressing reflection, but coupling a quantum model of a barrier to an environment, Model II, is effective.
These claims will be demonstrated in detail in the following sections.

We being in Section 2 with an analysis of the timescales involved in Models I and II. This analysis alone does in fact give significant evidence for our claims. Indeed, most of the physical understanding of our results is contained in Sections 1 and 2 of this paper and Sections 3,4 and 5 are devoted to confirming this physical understanding in precise technical detail for specific models.

In Section 3, we analyze Model I in the density matrix picture using lowest order perturbation theory to compute the scattering amplitude in the presence of an environment, thus generalizing the usual Born theory result. We find that reflection of not affected very much for the case $ L \propto \hat x$, but it is suppressed in the (unphysical) case $ L \propto \hat p $.

In Section 4, we analyze Model I in a very different way using the QSD approach. This gives us significant further insight into why the $L \propto \hat x $ case does not work very well, but the $L \propto \hat p $ case does. It also gives a nice visual
picture of the process whereby the reflecting wave packet is suppressed by the environment.

Sections 3 and 4 concentrate on the case of a real potential, but their conclusions apply
straightforwardly to the complex potential case and the Zeno effect. This is described in more detail in another paper \cite{BeHa2}.

In Section 5, we analyze Model II using lowest order perturbation theory to compute the scattering amplitude. We confirm that reflection is suppressed in this case, for suitable choice of parameters.

We summarize and conclude in Section 6.

\section{Timescale Analysis}

Much can be learned about the two models we are interested in by a simple analysis
of the timescales involved. This will be confirmed in the more detailed analytic
calculations of later sections.

\subsection{Timescales for Model I}

We suppose that the initial state is a spatially broad wave packet
\beq
\psi (x) = \frac{1} { (2 \pi \sigma^2)^{1/4} }\exp \left( - \frac{ (x-\bar x)^2 }{ 4 \sigma^2 } + \ih \bar p x \right),
\label{psi}
\eeq
with $\bar p>0$ (or perhaps a QSD localized state of the form Eq.(\ref{loc})).
There are two timescales of particular interest in relation to this packet.
One is the Zeno time, the time it takes the packet to traverse its own width,
\beq
t_z = \frac { m \sigma} {\bar p}.
\eeq
This is an essentially classical time.
The second we shall call the energy time and is the time it takes for the packet to traverse its own wavelength
\beq
t_E = \frac {\hbar} {E},
\eeq
where $E = \bar p^2 / 2m$. This is a quantum-mechanical time and it is the one most relevant to scattering processes. If the packet is reasonably well peaked in momentum space, $ \bar p \gg \hbar / \sigma$, then
\beq
t_E \ll t_z.
\eeq

For a packet scattering off a barrier of height $V_0$, we are interested in the case in which $E>V_0$, so that there is transmission and reflection. The reflection will be significant if $E $ is close to $V_0$, or in terms of the energy time
\beq
t_E \approx \frac {\hbar} {V_0},
\eeq
which indicates that $\hbar / V_0$ is the timescale on which the barrier acts. In the Zeno case,
recall that this corresponds to $ \hbar / V_0 = \eps$, the time spacing between projections. (See Refs.\cite{HaYe1,HaYe3} for similar timescale analyses.)

To get a good visual idea of the scattering process in the unitary case, we show in Fig.(1) the position and momentum space wave functions for a wave packet hitting a simple Gaussian potential as a sequence of snapshots. The whole sequence of snapshots is
taken on a total timescale $t_z$. Note that it takes some time for a clear reflected
wave packet to form. We will find in what follows that for reflection to be suppressed
decoherence has to be sufficiently fast that
the reflected packet is suppressed before it properly forms. Once it is formed, decoherence
will suppress interference between transmitted and reflected packets but does not
obviously suppress the probability of reflection.

\begin{figure}[h]
        \begin{center}
                \includegraphics[width=16cm]{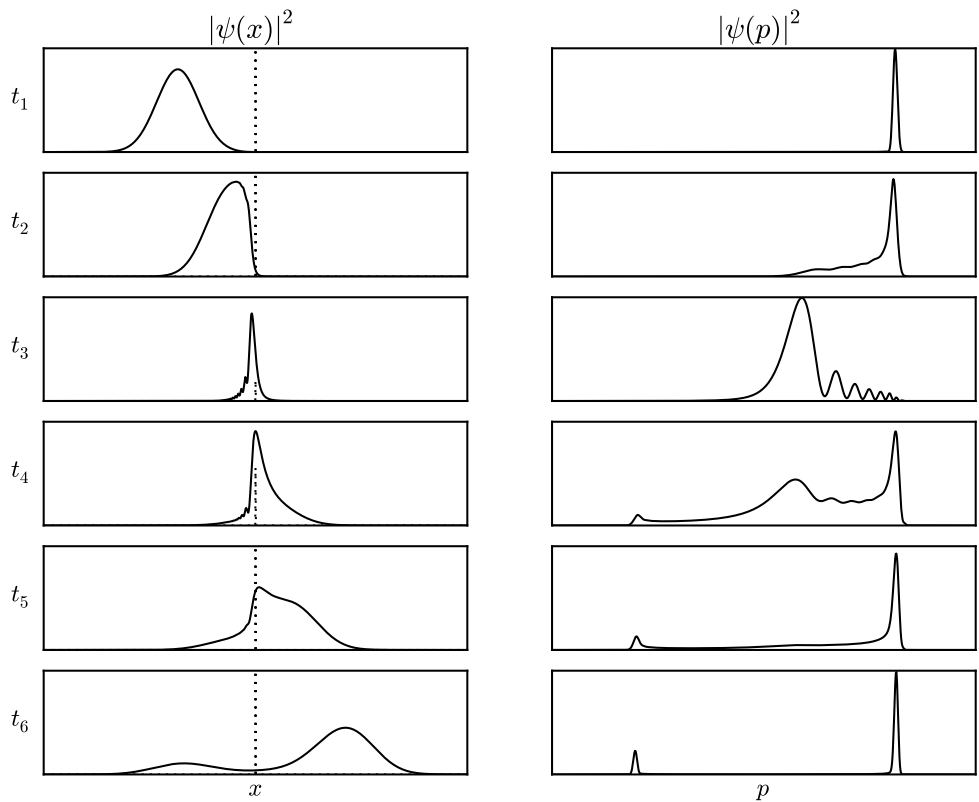}
        \caption{The position (left) and momentum (right) probabilities as a sequence of snapshots for an incoming wave packet incident on a Gaussian barrier. The total time duration of the whole sequence is of order $t_z$. The timespacing between snapshots is uneven and is chosen to best show how the probabilities change. In the final
        frames at time $t_6$, one can see in the position probability that the reflected and transmitted packets have formed and, correspondingly, in the momentum probability there are two clear peaks in the momentum, at
        $p = \bar p $ and $p = - \bar p $.}
        \end{center}
\end{figure}

Consider now the timescales that arise in the presence of an environment, with evolution described by the master equation Eq.(\ref{lin}) in the case $L \propto \hat x $, which we write out explicitly
in configuration space:
\begin{align}
\frac{\partial \rho}{\partial t} = \frac{i \hbar }{2m}\left(\frac{\partial^2 \rho}{\partial x^2} - \frac{\partial^2 \rho}{\partial y^2}\right) -  \ih \left(V(x)-V(y)\right)\rho
-\frac{D}{\hbar^2} (x-y)^2\rho.
\label{master1}
\end{align}
The most commonly discussed timescale associated with this equation is the decoherence time
\beq
t_d = \frac { \hbar^2} { D \ell^2},
\label{td}
\eeq
where $\ell$ is a lengthscale, typically chosen from the initial state. It is a particularly appropriate timescale when the initial state is a superposition of two spatially localized states a distance $\ell$ apart and is the timescale on which $ {\rm Tr} \rho^2 $ decreases from its initial
value of $1$ (for a pure initial state).
If $\ell$ is of macroscopic dimensions, the decoherence time is extraordinarily short and interferences are suppressed extremely effectively \cite{Zur2}.
In our case we have just a single incoming wave packet whose only relevant lengthscale is $\sigma$.
This could be quite large and hence $t_d$ very small. However, the decoherence time is related to position space diagonalization of the density matrix, and we will find that reflection is in fact strongly suppressed by momentum space diagonalization, which proceeds on a different timescale, discussed below.
The usual decoherence time is not therefore a key time scale in this model.

There is another significant timescale associated with Eq.(\ref{master1}), which is the localization time,
\beq
t_{loc} = \left( \frac {m \hbar} { D} \right)^{1/2},
\label{tloc2}
\eeq
defined in Eq.(\ref{tloc}) (repeated here for convenience).
It arises in QSD, as outlined above, but is also the timescale on which the Wigner function becomes positive under Eq.(\ref{Wig1}) \cite{DiKi}. This timescale is typically a lot longer than $t_d$ and turns out to be the key one for our scattering problem.

As stated earlier, we will analyze the problem in both the density matrix and QSD picture.
In the QSD picture, an initial state of the form Eq.(\ref{psi}) will evolve into a stochastic ensemble of localized states of the form Eq.(\ref{loc}), which will typically have
$\sigma_q \ll \sigma $ (since we assumed a broad initial wave packet). It is therefore natural in the QSD approach to explore initial states which are already in the localized form Eq.(\ref{loc}).
These localized states have a much shorter Zeno time,
\beq
t_z^{qsd} = \frac { m \sigma_q} {\bar p},
\eeq
where $\sigma_q$ is given by Eq.(\ref{widths}).

There are now three key timescales in the QSD analysis: $t_E$, $ t_z^{qsd}$ and $t_{loc}$.
Furthermore, we can fix the relationship between these scales by examining the requirement that the fluctuations produced by the environment do not affect the integrity of the wave packet. For the fluctuations in momentum to be acceptably small, in the localized packets of the QSD approach, we require that
\beq
\sigma_p \ll \bar p,
\eeq
where $\sigma_p$ is defined in Eq.(\ref{widths}). It is then easy to show that
\beq
\frac{t_E} {t_z^{qsd}} = \frac {\sigma_p} {\bar p},
\eeq
and
\beq
\frac{t_z^{qsd} } { t_{loc} } = \frac {\sigma_p} {\bar p}.
\eeq
The requirement of small fluctuations therefore leads to the important inequalities
\beq
t_E \ll t_z^{qsd} \ll t_{loc}.
\label{ineq}
\eeq

The inequalities Eq.(\ref{ineq}) now allow us to understand why the usual environment with $L \propto \hat x$ cannot suppress reflection without large fluctuations. For reflection to be suppressed, we would expect that the environment would have to act on a time scale much faster than
the potential acts, which means $ t_{loc} \ll \hbar / V_0$. But reflection is only significant when
$ \hbar / V_0 $ is of order $ t_E$. This means that $ t_{loc} \ll t_E$ which is not compatible with the small fluctuations requirement expressed in Eq.(\ref{ineq}).

The density matrix analysis, described in more detail below, leads to another timescale which reinforces the above conclusions. The master equation Eq.(\ref{master1}) produces diagonalization of the density matrix in position space on the timescale Eq.(\ref{td}). In the scattering calculations below, it is natural to look at this process in momentum space and we find that it produces diagonalization of the density matrix in momentum space, on a timescale
\beq
t_d^p = \left( \frac {m^2 \hbar^2 }{ D \bar p^2} \right)^{1/3},
\label{tmomdeco}
\eeq
where $p$ is the momentum scale of the incoming state.
It is easily shown that
\bea
\frac {t_z^{qsd} } {t_d^p} &=& \left( \frac {\sigma_p} {\bar p} \right)^{1/3},
\\
\frac {t_d^p} {t_{loc} } &=& \left( \frac {\sigma_p} {\bar p} \right)^{2/3}.
\eea
Combined with the earlier inequalities Eq.(\ref{ineq}) the requirement of small fluctuations now is
\beq
t_E \ll t_z^{qsd} \ll t_d^p \ll t_{loc}.
\label{ineq2}
\eeq
The important relation is the inequality between $t_E$ and $t_d^p$.

Consider now what is required for reflection to be suppressed. On purely dimensional
grounds from the timescales involved, one would expect that this requires momentum decoherence to be sufficiently fast that $ t_d^p < \hbar / V_0$. However, since $\hbar / V_0 \sim t_E $ this is incompatible with small fluctuations, Eq.(\ref{ineq2}).

The role of the momentum decoherence time can be confirmed more concretely by inspecting the Wigner equation, Eq.(\ref{Wig1}).
Reflection will be suppressed if the quantum terms in Eq.(\ref{Wig1}) are much smaller
than the remaining terms, since these remaining terms just describe classical stochastic
motion in a potential. We expect that these quantum terms to be smaller due to the spreading of the Wigner function produced by the diffusive term.
Taking the first of the quantum terms, this means we require
\beq
\left| V'(x) \frac { \partial W} { \partial p } \right |
\gg \left| \hbar^2 V'''(x) \frac {\partial^3 W }{ \partial p^3} \right|.
\label{Wigineq}
\eeq
The key scales in the problem are the momentum scale $ \bar p$ and the
spatial width at time $t$ of the state, $\sigma_t$. By estimating the order of magnitude
of each term, we see that
Eq.(\ref{Wigineq}) is the requirement that
\beq
\bar p^2 \sigma_t^2 \gg \hbar^2.
\label{Wigineq2}
\eeq
It is straightforward to show from Eq.(\ref{Wig1}) that $\sigma_t^2 \sim Dt^3/ m^2
$ for large $t$ and as a result Eq.(\ref{Wigineq2}) simply becomes the requirement
that $ t \gg t_d^p$. This indicates that the momentum decoherence time is indeed
a key timescale
for the suppression of reflection.

It is enlightening to consider the situation in which
the interaction with the environment is described by a master equation Eq.(\ref{lin})
in which $ L = D_p^{1/2} \hat p  $ considered in Ref.\cite{ChTe}.
This sheds some light on the process of suppression
of reflection although it is not a master equation that is obviously realized in any
physical relevant situations. This master equation produces diagonalization
of the momentum space density matrix on a timescale
\beq
t_p = \frac {1  } {D_p \bar p^2 },
\eeq
where $\bar p$ is the natural momentum scale of the problem so is taken to be the incoming
momentum. On general grounds, we would expect that reflection is significantly suppressed if this decoherence
process proceeds on a timescale shorter than the timescale $ t_E $ of the scattering
process, and we easily see that $t_p \ll t_E $ if
\beq
m \hbar  D_p \gg 1.
\label{cond}
\eeq
Now consider the issue of fluctuations generated by the environment in this case.
It is easy to show that, for the above choice of Lindblad operator,
the momentum diffusion term in
the Wigner equation Eq.(\ref{Wig1}) is replaced by a term of the form $ \hbar^2 D_p \partial^2
W / \partial x^2 $. This means there are envionmentally induced fluctuations in position
but not in momentum. There is therefore no problem with fluctuations in this case
and reflection is easily suppressed as long as $D_p $ is sufficiently large
that Eq.(\ref{cond}) holds. This simple model indicates that reflection is a momentum superposition effect, since it is significantly suppressed by momentum decoherence, but as stated it
is an unphysical model.

\subsection{Timescales for Model II}

We now give a qualitative analysis of Model II, the case of scattering off a classicalized target particle, described by Eq.(\ref{lin2}). Since the environment couples to the target and not to the light particle, the timescales in Model I referring to the action of the environment on the light particle play no role, although the energy
time $t_E $ and Zeno time $t_z$ are still relevant. Also, there are a number of new
timescales relating to the massive target particle. We denote all these by $T$, with
subscripts, to avoid confusion with the light particle timescales.

We suppose the target particle is in an initial Gaussian state with zero average position and momentum and spatial width $\Sigma$ and the incoming light particle has momentum $\bar p$.
This transfers momentum of this order to the target particle, which then has an associated Zeno time
\beq
T_z = \frac { M \Sigma} {\bar p},
\eeq
which is the time it takes to move a distance equal to its own width. We can also define a spatial decoherence time
\beq
T_d = \frac {\hbar^2 } {D \Sigma^2},
\eeq
and a localization time
\beq
T_{loc} = \left( \frac { M \hbar } { D} \right)^{1/2}.
\eeq
What will also be important is the size of the thermal fluctuations produced by the environment. The momentum fluctuations grow like
\beq
(\Delta P)^2_t  = \Sigma_p^2 + 2 D t,
\eeq
where $\Sigma_p$ is the initial momentum width which gives the timescale
\beq
T_f =  \frac{\Sigma_p^2 } {D},
\eeq
for the growth of momentum fluctuations. However,
we will suppose that $ \Sigma_p \sim \hbar / \Sigma $ which means that
this timescale is in fact the same as the decoherence time,
\beq
T_f = T_d.
\eeq
We can also define a momentum decoherence time
\beq
T_d^p = \left( \frac {M^2 \hbar^2 }{ D \bar p^2} \right)^{1/3},
\eeq
but note that this can be rewritten as
\beq
T_d^p = \left( \frac {M} {m} \right)^{1/3} T_{loc}^{2/3} t_E^{1/3}.
\eeq

Further relations between these timescales can be obtained by making the simplifying
assumption that the initial Gaussian of the target is a QSD localized state of the
type discussed in Section 1(D). This means that the spatial width of the state is
given by
\beq
\Sigma = \left( \frac { \hbar^3 } { 8 M D} \right)^{1/4}.
\eeq
This is a natural assumption. It corresponds to the idea that the target has been
interacting with the environment for a few localization times and has therefore settled
down in a mixed state of the form Eq.(\ref{rholoc}) and we take one of its diagonal
elements as the initial state. This value of $\Sigma$ leads to the relations
\bea
T_d &=& T_{loc},
\\
T_z &=& \left( \frac {M}{m} \right)^{1/2}T_{loc}^{1/2} t_E^{1/2}.
\eea
We therefore now have $T_f = T_d = T_{loc}$ and $T_z$, $T_d^p$
are expressed in terms of $T_{loc}$ and $t_E$.

We can now ask, purely on the grounds of timescales, under what conditions we would expect reflection to be suppressed in this model. Generally, we would expect that
this will be the case if the five timescales $ T_z, T_d, T_f, T_d^p, T_{loc}$ are
much less than the some characteristic timescale of reflection, which could involve
$t_E$, $t_z$ and perhaps also powers of $ m / M$.

Some further input is required to fix the relationship between timescales more
precisely. To this end, we consider the relationships between the initial momenta
$ \bar P, \bar p $ and final momenta $ P,p$ required by energy and momentum conservation.
One easily finds
\bea
P &=& \frac {M-m} {M+m} \bar P + \frac {2 M} { M+m } \bar p,
\\
p &=& \frac {2m} {M+m} \bar P - \frac {M-m} {M+m} \bar p.
\eea
These relations will hold exactly in the unitary case in quantum theory and will
still hold in some approximate sense when the environment is present.
We assume that $ m \ll M$ and to first order in $m/M$ we obtain
\beq
p = - \bar p + 2 \frac {m} {M} P,
\eeq
or in terms of velocities $u = p/m$ and $U = P/ M$, we have
\beq
u = - \bar u + 2 \bar U.
\eeq
As discussed, we anticipate that reflection is suppressed as a result of the fluctuations
of the target being transferred to the light particle, and this relation shows it
is the fluctuations in {\it velocity} that are important. This suggests that the
conditions under which reflection is suppressed will depend on the velocity of the light
particle, not its momentum (i.e. will not depend on its mass).

The above argument leads us to suggest that conditions for the suppression of reflection
can contain $t_E$ only in the combination $m t_E$. So one such simple condition would be
\beq
T_{loc} \ll \frac {m} {M} t_E,
\label{res2}
\eeq
which also of course implies that $T_f $ and $T_d$ have the same upper bound.
This condition also implies the two conditions
\bea
T_z & \ll & t_E,
\\
T_d^p & \ll & \left( \frac {m}{M} \right)^{1/3} t_E.
\label{tmom1}
\eea
Note that Eq.(\ref{res2}) may be re-written,
\beq
\frac {p} {m}  \ll \frac {\Sigma_p} {M},
\label{pres}
\eeq
where, recall, $\Sigma_p \sim ( M \hbar D )^{1/4} $, the momentum width of the
target particle initial state, which means that the velocity of the incoming
particle must be smaller than the velocity fluctuations of the target,
as anticipated. Further conditions of this general form will arise in
the more detailed calculation of Model II later on.

These (possibly large) fluctuations in the target particle
are not a problem in this model, since there are two different particles involved,
and the model does not suffer from the fluctuation problem of Model I.
However, we should check that the restriction Eq.(\ref{pres}) is compatible with
our requirement of small fluctuations for the light particle. We easily find
that Eq.(\ref{pres}) and $ \bar p \gg \hbar / \sigma $ are satisfied if
\beq
\frac { M \Sigma} {\bar p}  \ll \frac {m \sigma } {\bar p},
\eeq
that is, the Zeno time of the target is much shorter than the Zeno time of
the incoming packet. This can be satisfied if the incoming packet is sufficiently
broad, which we have always assumed.

Note that physically speaking, the large velocity fluctuations of the target will require some sort of containment to stop them from growing indefinitely, which would be physically unrealistic. We therefore imagine that there is some sort of containment, such as a harmonic oscillator potential, but that the whole process contemplated here takes place on a time scale short compared to the time it takes the system to experience the effects of its confining potential.

The more detailed analysis given below of Model II will in fact reveal a more elaborate picture
of the suppression of reflection. We will find that there are some circumstances in which the environment is not in fact needed and that fluctuations in the initial state of the target particle are in fact sufficient to suppress reflection. However, an environment is required in the most general case. Furthermore, the general idea that
large velocity fluctuations of the target cause the suppression of reflection hold in all cases.

\section{Perturbative analysis of Model I. Environment coupled to incoming particle}
\label{env1}
In this section we analyze Model I by solving the Lindblad equation Eq.(\ref{lin}) for an incoming wave packet initial state to lowest order in perturbation theory in $V_0$, thus obtaining the probability for reflection. This
generalizes the Born result Eq.(\ref{Uref}) to the situation in which a decohering environment is present. We focus on the case of the usual master equation, with Lindblad operator $L \propto \hat x $ but for comparison we also consider the case $L \propto \hat p $.
The probability for the transmitted packet is trivial for small $V_0$ -- it is just
the unperturbed result plus a correction of order $V_0^2$ and we do not compute this
explicitly. In what follows we will
concentrate entirely on the reflected part.

\subsection{The Case $L \propto \hat x$ }

In the case $L = \hat x$ the master equation is given in position space by Eq.(\ref{master1}).
We solve this perturbatively for a wave packet initial state. The solution is obtained
using the density matrix propagator for the case $V=0$,
\begin{align}
J(x,y,t | x',y',t')  =& \frac{m}{2\pi\hbar (t-t')} \exp\left( \frac{im}{2\hbar (t-t')}\left[(x-x')^2-(y-y')^2\right]\right)
                                \nonumber\\
                                &\times  \exp\left(- \frac{D(t-t')}{3\hbar^2}\left[ (x-y)^2 +(x-y)(x'-y')+(x'-y')^2  \right]\right),
\end{align}
(see for example Refs.\cite{CaLe,AnHa}).
It is also useful to have the momentum space form of this propagator which is
\begin{align}
J(p,q,t | p',q',t')  = & \frac{1}{\sqrt{4\pi D(t-t')}}  \delta(p-q-p'+q')\nonumber\\
                                &\times \exp\left(-\frac{i(t-t')}{4m\hbar}(p^2-q^2+p'^2-q'^2)\right)
                                \nonumber\\
                                &\times \exp\left( - \frac{1}{4D(t-t')}(p-p')^2 - \frac{D(t-t')^3}{12m^2\hbar^2}(p-q)^2\right).
\end{align}

The scattering process is governed by the various timescales discussed in Section
2 and recall that in order to avoid large fluctuations, these timescales must lie
in the regime defined by
the inequalities Eq.(\ref{ineq2}). This means that in this regime we
can make the following useful approximation \cite{Yeapp}
\begin{align}
\left. \frac{1}{\sqrt{4\pi Dt}}\exp\left( - \frac{1}{4Dt} (p-p')^2\right)\right|_{t\lesssim t_z}\sim \delta(p-p').
\label{Yearsley}
\end{align}
We expect this to hold as long as the momentum scale $\bar p $ satisfies $\bar p^2 \gg D t_z $.
This condition means that in the course of the experiment the diffusive momentum spreading due to the environment does not destroy the integrity of the wavepacket, in keeping with the general ideas described in Section 2.
We can then approximate the momentum space density matrix propagator as
\begin{align}
J(p,q,t | p',q',t')  & \simeq \;  \delta(p-p')\delta(q-q')
\nonumber \\
                        & \times \exp\left(-\frac{i(t-t')}{2m\hbar}(p^2-q^2) - \frac{D(t-t')^3}{12m^2\hbar^2}(p-q)^2 \right).
\end{align}
An approximation of the form Eq.(\ref{Yearsley}) is not possible for the final term in the exponential.

A completely standard perturbation analysis of Eq.(\ref{master1}) gives the
second order solution
\begin{align}
\rho_t(p,q) = & \frac{1}{2\pi\hbar}\int_0^t dt' \int_0^{t'} dt'' \int  \left[\prod_{i = 0}^4 {dp_i}{dq_i} \right] J(p,q,t|p_4,q_4,t') \nonumber\\
&\times  \frac{ i}{ \hbar}\left(V(p_4-p_3)\delta(q_4-q_3) - V(q_4-q_3)\delta(p_4-p_3)\right)\nonumber\\
&\times  J(p_3,q_3,t'|p_2,q_2,t'') \nonumber\\
&\times \frac{ i}{ \hbar}\left(V(p_2-p_1)\delta(q_2-q_1) - V(q_2-q_1)\delta(p_2-p_1)\right)\nonumber\\
&\times  J(p_1,q_1,t''|p_0,q_0,0) \rho_0(p_0,q_0).
\end{align}
There are zeroth and first order terms also in the solution but this expression is the lowest
order contribution to reflection. Note that since the system is coupled to an environment
which acts at all times,
we cannot take the initial state at $t\rightarrow -\infty$ as this would generate an infinite momentum dispersion by the time the wave packet reached the origin.

The potential function in momentum space is given by
\begin{align}
V(p) = \frac{1}{\sqrt{2\pi\hbar}}\int dx \exp\left(-\frac{i}{\hbar}px\right)V(x).
\end{align}
For example, the Gaussian potential in momentum space is given by
\begin{align}
V(p) = V_0\frac{1}{\sqrt{2 \pi \hbar}}\exp\left(-\frac{a^2}{2\hbar^2}p^2 \right).
\label{Gauss}
\end{align}
Only two of the four possible combinations of the potential terms contribute to the reflected norm. The others result in an overall factor of $\delta(p-p_0)$ or $\delta(q-q_0)$. After integrating out all $\delta$-functions the we find that the reflected norm is
\begin{align}
\rho_t^{\rm ref} (p,p)  = &  \frac{1}{2\pi\hbar^3}\int_0^t ds \int_0^{t-s} du \int dp_0 dq_0
 V(p_0-p)V(p-q_0)\nonumber\\
&\times \left[ \exp\left(-\frac{is}{2m\hbar}(p^2-q_0^2) - \frac{Ds^3}{12m^2\hbar^2}(p-q_0)^2\right)\right.\nonumber\\
&\quad\quad \left.+\exp\left(-\frac{is}{2m\hbar}(p_0^2-p^2) - \frac{Ds^3}{12m^2\hbar^2}(p_0-p)^2\right)
\right]\nonumber\\
&\times \exp\left(-\frac{iu}{2m\hbar}(p_0^2-q_0^2) - \frac{Du^3}{12m^2\hbar^2}(p_0-q_0)^2\right)\rho_0(p_0,q_0).
\end{align}
where $u=t^{\pp}$ and $s = t' - t^{\pp}$.
The remaining momentum integrals can be performed once we assume a form for the initial density matrix. This we take to be a very broad wave packet which we approximate
as a pure plane wave
\begin{align}
\rho_0(p_0,q_0) &=  \sqrt{\frac{2}{\pi}}\frac{\sigma}{\hbar}\exp\left(-\frac{\sigma^2}{\hbar^2}\left[(p_0-\bar{p})^2
+(q_0-\bar{p})^2\right]-\frac{i}{\hbar}\bar{x} (p_0- q_0)\right)
\label{initialrho}\\
&\simeq \frac{\sqrt{2\pi}\hbar}{\sigma}\delta(p_0 - \bar{p})\delta(q_0 - \bar{p}),
\end{align}
where $\bar x = ( \pi \sigma^2 / 2)^{1/2}$.
This results in
\begin{align}
\rho_\tau^{\rm ref} (p,p) = &\frac{2m}{\hbar^2\bar{p}} V^2(p-\bar{p})
\nonumber\\& \times
\int_0^\tau ds \left(1-\frac{s}{\tau} \right)\cos\left(\frac{s}{2m\hbar}(p^2-\bar{p}^2)\right) \exp\left(- \frac{Ds^3}{12m^2\hbar^2}(p-\bar{p})^2\right),
\end{align}
where we have introduced the timescale $ \tau = 2 m \bar x / \bar p  $ (which is of order $t_z$).
Taking the limit $\tau\rightarrow \infty$ gives
\begin{align}
\rho_\infty^{\rm ref} (p,p) = \frac{2 m}{\hbar^2\bar{p}} V^2(p-\bar{p})
\int_0^\infty ds \cos\left(\frac{s}{2m\hbar}(p^2-\bar{p}^2)\right) \exp\left(- \frac{Ds^3}{12m^2\hbar^2}(p-\bar{p})^2\right).
\label{Airy}
\end{align}

Eq.(\ref{Airy}) is the desired result and reduces to the Born approximation result
Eq.(\ref{Uref}) in the case $D=0$. It can be evaluated for non-zero $D$ but this is not in fact necessary to deduce the result we are interested in. The effects of
the environment on this scattering probability are described by the exponential
term and its effect is easily estimated by considering the timescales arising
in the integral. We are interested in the case where the final momentum $p$
is of order $ - \bar p$. It is then easily seen that the cosine term contributes to the integral on a time
scale $ s \sim t_E $ but the exponential term only contributes for $ s  > t_d^p$,
the momentum decoherence timescale. The requirement of small fluctuations
Eq.(\ref{ineq2}) means that $ t_E \ll t_d^p$, which implies that the
exponential term makes negligible contribution to the value of the integral.

This calculation therefore shows that, in the regime of small fluctuations, the only regime where the results
make any physical sense, the environment has negligible effect on the usual
reflection probability so reflection is not suppressed. This confirms part of the
general story outlined in the Introduction.

On general grounds,
if we allowed large fluctuations, we would then expect that
the reflected momentum distribution will no longer be concentrated around $  p \approx -\bar p $ for sufficiently large $D$ but will be very spread out, and subsequently suppressed (by the potential term and prefactor), as described in the Introduction. Eq.(\ref{Airy}) indicates this behaviour but this is misleading since
the approximation Eq.(\ref{Yearsley}) used to derive it is no longer valid for large
fluctuations.
To prove this in the large fluctuation regime, we would therefore have to go beyond
the approximation Eq.(\ref{Yearsley}), which would lead to a much more complicated
integral. We have not carried this out explicitly, but we would expect the result
to be very similar in flavour to Eq.(\ref{Airy}).

\subsection{The Case $L \propto \hat p$}

We now for comparison repeat the above calculation in the case $L = D_p^{1/2} \hat p $.
The momentum space density matrix propagator in this case (for $V=0$) is
\begin{align}
J(p,q,t|p',q',t') =& \delta(p-p')\delta(q-q')
\nonumber \\
\times & \exp\left(-\frac{i(t-t')}{2m\hbar}(p^2-q^2) - D_p(t-t')(p-q)^2\right).
\end{align}
Following the same second order perturbative calculation carried out above we obtain
\begin{align}
\rho_\infty^{\rm ref} (p,p) = \frac{2 m}{\hbar^2\bar{p}} V^2(p-\bar{p})
\int_0^\infty ds \cos\left(\frac{s}{2m\hbar}(p^2-\bar{p}^2)\right) \exp\left(- {D_p
s}(p-\bar{p})^2\right).
\end{align}
The integral can be performed with result
\begin{align}
\rho_\infty^{\rm ref} (p,p) = \frac{2\pi m^2}{\hbar\bar{p}^2}& V^2(p-\bar{p})
\frac{ 4m\hbar D_p \bar{p}(p-\bar{p})^2}{\pi[(p^2-\bar{p}^2)^2+(2m\hbar D_p)^2(p-\bar{p})^4 ]}.
\label{refnorm1}
\end{align}

\begin{figure}[h]
        \begin{center}
                \includegraphics[width=14cm]{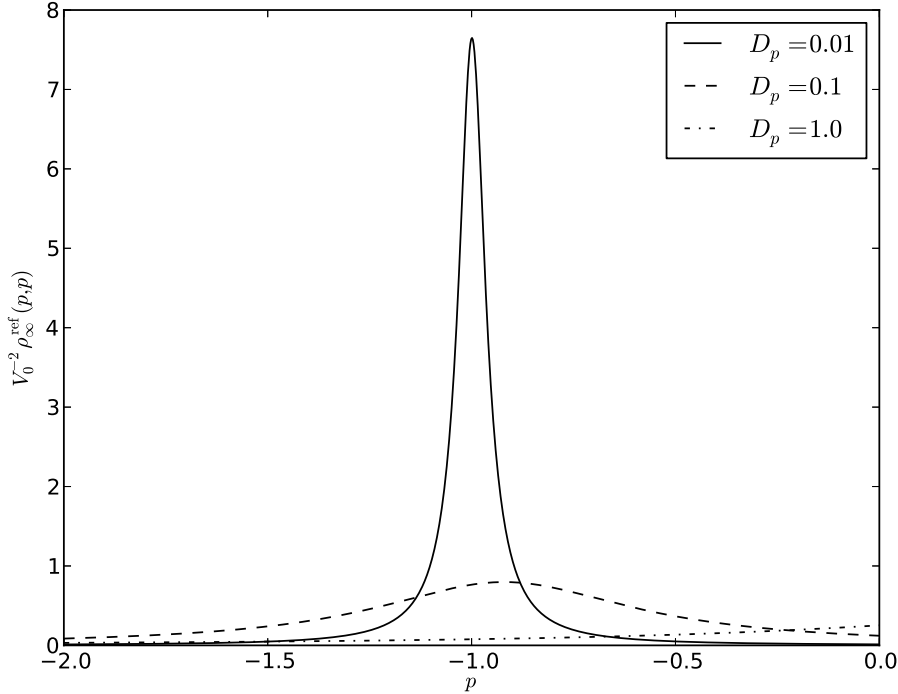}
        \caption{Plot showing the reflected probability density $\rho_{\infty}^{\rm ref}(p,p)$ as a function of momentum $p$ (see Eq.(\ref{refnorm1})) in the case where $L\propto p$. The potential barrier is a Gaussian with associated length scale $a$. Plots are shown for a range of values of the diffusion parameter $D_p$.
        For small $D_p$ the reflected probability density tends to a delta function about $p = -\bar{p}$. As $D_p$ increases the reflected peak spreads and flattens. By around $D_p=1$ the reflected peak is seen to spread well into the transmitted region ($p>0$). Units are chosen such that $m = \bar{p} = \hbar = 1$ and we choose $a=0.1$.}
\end{center}
\end{figure}

\begin{figure}[h]
        \begin{center}
                \includegraphics[width=13cm]{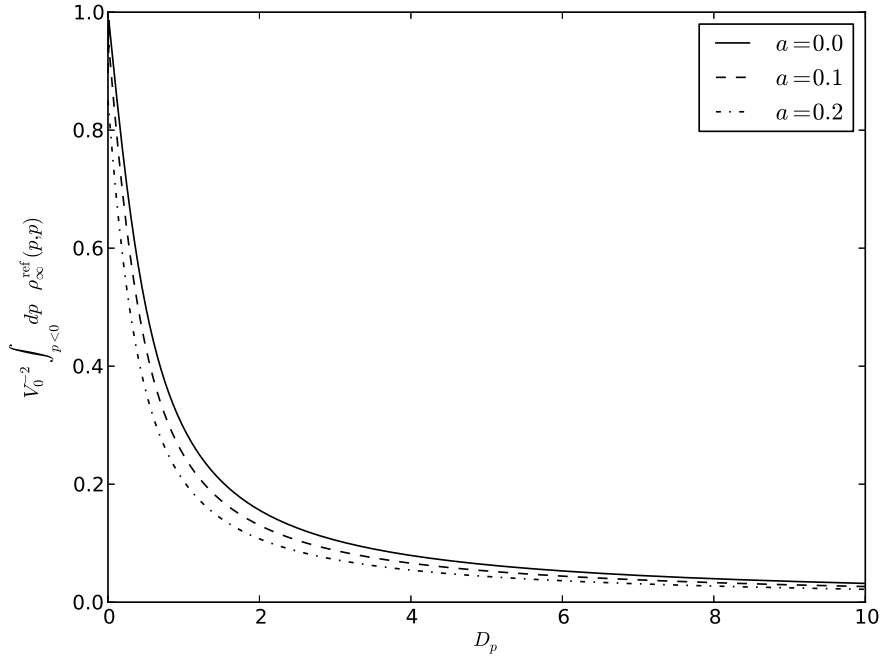}
        \caption{Plot showing the total reflected probability as a function of $D_p$ in the case where $L\propto p$. The potential barrier is a Gaussian with associated length scale $a$ as in Fig.(2).
        It is clear that $D_p\sim 1$ sets a scale at which reflection in suppressed. Comparing with Fig.(2) we see that this is also the point at which the reflected peak spreads out beyond the $p=0$ boundary. Plots are shown for increasing values of $a$. Larger values of $a$ have the effect of decreasing the probability of reflection generally. Units are chosen such that $m = \bar{p} = \hbar = 1$.}
\end{center}
\end{figure}

In the limit $D_p\rightarrow 0$ the last term in this expression tends to the delta function $\delta(p+\bar{p})$ giving agreement with equation (\ref{Uref}) upon using the Gaussian potential. Otherwise, the result is a smeared out
distribution concentrated around $ p = - \bar p $, with the degree of smearing
depending on the quantity $ m \hbar  D_p $.

A plot of the reflected norm with the potential given by (\ref{Gauss}) is shown in Fig.(2). We see that it is very spread out for $ m \hbar D_p \gg 1 $ in agreement
with the general analysis of Section 2, where we deduced this condition from the
requirement that $t_p \ll t_E$ (where $t_p$ is the momentum decoherence time for this case).
As discussed in the Introduction, this spreading pushes the reflected momentum into
the positive momentum regime or into the very negative regime where it is suppressed
and it is in this way that the total reflected momentum is suppressed. The total reflected norm
as a function of $D_p$ is shown in Fig.(3), where we see that it is indeed suppressed for large $D_p$.
As argued already, there is no fluctuation problem in this model since the
Lindblad operators $L \propto \hat p $ do not produce momentum fluctuations, only
momentum decoherence. However, as stated, this is an unphysical model, but it supports
the physical interpretation that reflection is suppressed by momentum decoherence.

The possible suppression of reflection with a Lindblad master equation with
$L \propto \hat p $ was previously considered in Ref.\cite{ChTe}. A general
expression for the reflection probability was given and it was speculated that it
gave suppression of reflection although it was not evaluated explicitly as here. Furthermore, the more physical case $ L \propto \hat x $ considered here and its
associated difficulty with large fluctuations
was not discussed.

\section{QSD analysis of Model I}

As discussed in the Introduction a useful alternative way to analyse Model I is using the QSD approach
in which the state evolves according to the stochastic differential equation Eq.(\ref{qsd1}).
We will solve the system for three different situations and confirm from a different perspective the results of Sections 2 and 3.

\subsection{Steady state solution for the case $ L \propto \hat x$}

For the case $L \propto \hat x$, Eq.(\ref{qsd1}) reads
\begin{align}
d|\psi\rangle = -\frac{i}{\hbar}(H_0+V)|\psi\rangle dt - \frac{D}{\hbar^2}(x-\langle x\rangle)^2|\psi\rangle dt + \frac{\sqrt{2D}}{\hbar}(x-\langle x\rangle)|\psi\rangle dB_t.
\label{QSD}
\end{align}
To begin we assume that the incoming wave packet is in the steady state
\begin{align}
\langle x | \psi\rangle = \frac{1}{\sqrt[4]{2\pi\sigma_q^2}}
\exp\left(-\frac{(1-i)}{4\sigma_q^2}(x - \langle x \rangle)^2 +\frac{i}{\hbar}\langle p \rangle x\right),
\end{align}
where, recall, $\sigma_q^2 = \sqrt{\hbar^3/8Dm}$. We further assume that the potential has a small perturbative effect about the steady state solution. This will require a sufficiently small potential (see below).

The approximate steady state assumption corresponds to the idea that the wave packet is behaving in an approximately classical fashion as it traverses the potential barrier. We aim to examine the conditions under which this is a valid approximation. To make the calculations easier we choose a step potential of the form $V(x) = V_0\theta (x)$. The Hamiltonian is $H_0 = p^2/2m$.

In general for an operator ${A}$ the process satisfied by its quantum expectation $\langle A \rangle = \langle\psi|A|\psi\rangle/\langle\psi |\psi\rangle$ is
\begin{align}
d\langle A \rangle = -\frac{i}{\hbar}\langle [A,(H_0+V)] \rangle dt - \frac{D}{\hbar^2}\langle [[A,x],x]\rangle dt +\frac{\sqrt{2D}}{\hbar}\langle Ax+xA-2A\langle x\rangle\rangle dB_t.
\end{align}
This equation enables us to write down several useful processes:
\begin{align}
d\langle x \rangle =& \frac{1}{m}\langle p \rangle dt + \frac{\sqrt{8D}}{\hbar}{\rm Var}(x)dB_t,
\label{xprocess}\\
d\langle p \rangle =& -V_0|\psi(0)|^2 dt + \frac{\sqrt{8D}}{\hbar}{\rm Cov}(x,p)dB_t,
\label{pprocess}
\end{align}
\begin{align}
d{\rm Var}(x)  = &  \frac{2}{m}{\rm Cov}(x,p)dt -\frac{8D}{\hbar^2}{\rm Var}^2(x)dt  \nonumber\\
& + \frac{\sqrt{8D}}{\hbar}{\rm Cov}(x,x^2) dB_t
	 -\frac{4\sqrt{2D}}{\hbar}\langle x\rangle {\rm Var}(x) dB_t,
\label{varx}\\
d{\rm Var}(p) = & -2mV_0J(0)dt +2V_0\langle p \rangle |\psi(0)|^2dt + 2D\left(1-\frac{4}{\hbar^2}{\rm Cov}^2(x,p)\right)dt \nonumber\\
&+ \frac{\sqrt{8D}}{\hbar}{\rm Cov}(x,p^2) dB_t - \frac{4\sqrt{2D}}{\hbar}\langle p\rangle {\rm Cov}(x,p) dB_t,
\label{varp}
\end{align}
\begin{align}
d{\rm Cov}(x,p)  = & \frac{1}{m}{\rm Var}(p)dt +V_0\langle x\rangle|\psi(0)|^2dt - \frac{8D}{\hbar^2}{\rm Var}(x){\rm Cov}(x,p)dt
\nonumber\\&+\frac{\sqrt{2D}}{\hbar}{\rm Cov}(xp+px,x)dB_t
-\frac{\sqrt{8D}}{\hbar}\langle x \rangle {\rm Cov}(x,p)dB_t
\nonumber\\&- \frac{\sqrt{8D}}{\hbar}\langle p\rangle {\rm Var}(x) dB_t,
\label{cov}
\end{align}
where ${\rm Var}(A) = \langle A^2\rangle - \langle A\rangle^2$, ${\rm Cov}(A,B) = \langle AB+BA\rangle/2 - \langle A\rangle\langle B\rangle$, $\psi(x) = \langle x|\psi\rangle$, and the current $J(x)$ is given by
\begin{align}
J(x) =\frac{\hbar}{2mi}\left(\psi^*(x)\frac{\partial\psi(x)}{\partial x}-\frac{\partial\psi^*(x)}{\partial x}\psi(x)\right).
\end{align}
For the steady state solution (when $V = 0$) we have ${\rm Var}(x) = \sigma_q^2$, ${\rm Var}(p) = \hbar^2/2\sigma_q^2$, and ${\rm Cov}(x,p) = \hbar/2$, and the processes describing each of these second order moments are static.

Now consider the steady state wave packet interacting with the potential. We have solved Eqs (\ref{xprocess}) and (\ref{pprocess}) using a combination of numerical and analytic methods,
assuming a steady state packet throughout. The calculation gives the classical result, i.e., pure transmission with the correct average momentum (plus stochastic fluctuations) when the energy is greater than the potential and pure reflection when the energy is less than the potential. Since we have assumed that the wave packet maintains its shape as it crosses the potential barrier this calculation indicates that the formation of quantum reflection must be related to deformations of the wave packet from its steady state form.

For the steady state solution, the potential terms in (\ref{varp}) are
\begin{align}
-2mV_0J(0)dt +2V_0\langle p \rangle |\psi(0)|^2dt
= \frac{\hbar V_0}{\sigma_q^2}\langle x\rangle |\psi(0)|^2 dt.
\end{align}
The combination $\langle x\rangle |\psi(0)|^2$ is at most ${\cal O}(1)$ for a Gaussian wave packet. We conclude that the variance in $p$ will grow at a rate of order $V_0/\hbar$. This can be thought of as characterising the growth of the reflected peak. The only way in which the environment can act to suppress this growth in ${\rm Var}(p)$ is by generating a covariance in $x$ and $p$ greater than the steady state value (see third term in (\ref{varp})).

The potential term in (\ref{cov}) also leads to growth of ${\rm Cov}(x,p)$ at a rate $V_0/\hbar$. This means that the variance in $p$ will no longer grow as a result of the interaction with the potential after a time scale  $t$ such that
\begin{align}
\frac{\hbar V_0}{\sigma_q^2}\langle x\rangle |\psi(0)|^2 +2D\left[1-\left(1+\frac{V_0}{\hbar}t\right)^2\right] = 0.
\end{align}
To lowest order in $V_0$ this implies
\begin{align}
t \sim \frac{\hbar^2}{D\sigma_q^2} \sim t_{loc}.
\end{align}
(This time scale is also of the same order as $t_d$ and $t_d^p$ for the steady state packet where the length and momentum scales are defined by the packet size.)
In other words the environment is effective at suppressing growth of $p$-variance on the localization time scale. However, as we have seen in earlier sections, the appropriate time scale associated with formation of reflection is $t_E$. We therefore require that $t_{loc}\ll t_E$ in order that the environment effectively suppresses reflection. As shown in Section 2 this is in contradiction with the requirement that the momentum fluctuations should be small.

\subsection{More general solution for the case $L \propto \hat x $}

We now consider the more general case in which the initial state is a more general wave packet, not of the steady state form. The localization process and suppression of reflection will now be governed by different timescales. These can be found from an examination of Eqs(\ref{varx})-(\ref{cov}). In what follows we ignore the potential terms.

From equation (\ref{varx}) it is evident that the effect of the environment is to reduce ${\rm Var}(x)$ on average so that the state becomes more localized. This occurs on a timescale of order $\hbar^2/D\sigma^2$ -  the decoherence timescale - where $\sigma$ is the width of the wave packet (not yet localized).

We can also demonstrate that the state has a tendency to localize in momentum space. Consider an initial state with $2{\rm Cov}(x,p)/\hbar = c \gg 1$. From equation~(\ref{varp}) we see that ${\rm Var}(p)$ will decrease on average on a timescale of order ${\rm Var}(p)/(Dc^2)$. However, now consider that ${\rm Cov}(x,p)$ is initially small such that $c\ll1$. Before ${\rm Var}(p)$ can decrease, the covariance must grow to a sufficient value. Suppose that this happens after a time $t$ in which case, from equation (\ref{cov}), $c\sim{\rm Var}(p) t/m\hbar$, so that the total timescale for the momentum to become localized is given by
\begin{align}
t_d^p \sim t + \frac{m^2\hbar^2}{D{\rm Var}(p)t^2}.
\end{align}
A simple way to determine $t$ is to optimise $t_d^p$ with respect to $t$. Doing this we find
\begin{align}
t_d^p \sim \left(\frac{m^2\hbar^2}{D{\rm Var}(p)}\right)^{1/3}.
\end{align}
During the process of quantum mechanical reflection we have ${\rm Var}(p)\sim \bar{p}^2$ and so we recover the momentum decoherence timescale. Note that choosing ${\rm Var}(p)\sim \bar{p}^2$ typically imposes the weakest available constraint on $t_d^p$. For suppression of reflection we would typically require the momentum decoherence time to be shorter than the timescales associated with reflection. Since $\bar{p}$ is the longest momentum scale available, by using it we are being generous in defining $t_d^p$ to be as small as possible. In fact it is likely that momentum decoherence would need to act whilst ${\rm Var}(p)$ is much smaller. However, since this choice is enough to demonstrate that suppression cannot happen without large momentum fluctuations, that is all we require. Indeed,
if momentum decoherence is to act effectively to suppress refection we require that $t_d^p\ll t_E$. Again this is in contradiction to the requirement of small momentum fluctuations discussed in earlier sections.

Another way to see this is as follows. From Eqs.(\ref{pprocess}) and (\ref{varp}) we find that $t_f =\bar{p}^2/D$ is the timescale on which fluctuations in momentum are on the scale $\bar{p}$. This can seen since from (\ref{varp}), the quantum variance in $p$ grows as $2D(1-c^2)t$ on average; from (\ref{pprocess}), the stochastic variance in $\langle p \rangle $ grows as $2Dc^2t$. In total the momentum fluctuations grow as $2Dt$ from which follows the fluctuation timescale. Now if $t_d^p\ll t_E$ for ${\rm Var}(p)\sim \bar{p}^2$ (the weakest way to impose this constraint) then it follows that $t_f\ll t_E$ and the momentum fluctuations will destroy the wave packet on a much shorter timescale than that associated with the reflection.

We thus see that for more general initial states, localization and suppression of reflection proceed on the timescale $t_d^p$ which is shorter than the localization timescale, but it is still not sufficiently short to avoid the large fluctuations problem. We thus confirm from a different angle the results of Sections 2 and 3.

\subsection{The case $L \propto \hat p $}

We also consider the case where the Lindblad operator $L \propto \hat p$. The density matrix for a free particle in this case is
\begin{align}
\frac{d\rho}{dt} = -\frac{i}{2m\hbar}(p^2-q^2)\rho - D_p (p-q)^2\rho.
\end{align}
This corresponds to a quantum state evolution of the form
\begin{align}
d|\psi\rangle = -\frac{i}{\hbar}(H_0+V)|\psi\rangle dt - {D_p}(p-\langle p\rangle)^2|\psi\rangle dt + \sqrt{2D_p}(p-\langle p\rangle)|\psi\rangle dB_t.
\label{QSD2}
\end{align}
The stochastic processes $\langle p \rangle $ and ${\rm Var}(p)$ can then be shown to satisfy
\begin{align}
d\langle p \rangle =& -V_0|\psi(0)|^2 dt + \sqrt{8D}{\rm Var}(p)dB_t,
\label{pprocess2}\\
d{\rm Var}(p)  =& -2mV_0J(0)dt +2V_0\langle p \rangle |\psi(0)|^2dt-8D_p{\rm Var}^2(p)dt
\nonumber\\&+ \sqrt{8D_p}{\rm Cov}(p^2,p)dB_t
-4\sqrt{2D_p}\langle p\rangle {\rm Var}(p)dB_t.
\label{varp2}
\end{align}
From the third term on the right-hand side of Eq.(\ref{varp2}) we see that the environment tends to localize the state of the particle in momentum space on a timescale $t_p = 1/(D_p{\rm Var}(p))$. This is the momentum decoherence time for this environment.

Meanwhile from (\ref{pprocess2}) we find that the stochastic variance of $\langle p \rangle$ grows as $8D{\rm Var}^2(p)t$ and from (\ref{varp2}) the quantum variance of $p$ decreases as $-8D{\rm Var}^2(p)t$ on average. These effects cancel each other out and there is no growth of fluctuations in $p$ on average.

So for $L\propto \hat p$ there is no problem with the condition that $t_p \ll t_E$. For ${\rm Var}(p)\sim \bar{p}^2$ this simply leads to the condition that $m \hbar  D_p \gg 1$. This can be met with no effect on momentum fluctuations as discussed in
Section 3. In the limit that ${\rm Var}(p)\rightarrow 0$ (plane wave) we see from (\ref{pprocess2}) that $\langle p \rangle$ exhibits no stochastic behaviour whilst from (\ref{varp2}) we see that the variance has no tendency to grow.
This is all fully consistent with the analogous calculations of Section 3.

\section{Perturbative analysis of Model II. Environment coupled to target}

We now consider Model II in which the potential term arises due to the presence of a heavy target particle which has the possibility of being coupled to an environment, but the incoming light particle is not coupled to the environment. This avoids the fluctuations problem of Model I.
We consider the two cases of with and without environment and within those cases, there is also the possibility of averaging over the final state of the target particle, or keeping it fixed, so there are four cases in total.

\subsection{Pertubative solution to the Master equation}

The Hamiltonian for this model is given by Eq.(\ref{ham}) and the master equation is given by Eq.(\ref{lin2}), which may be written explicitly as
\begin{align}
\frac{\partial \rho}{\partial t} = & \frac{i\hbar}{2m}\left(\frac{\partial^2 \rho}{\partial x^2} - \frac{\partial^2 \rho}{\partial y^2}\right)
        +\frac{i\hbar}{2M}\left(\frac{\partial^2 \rho}{\partial X^2}
- \frac{\partial^2 \rho}{\partial Y^2} \right) \nonumber\\
& - \frac{i}{\hbar}\left(V(x-X)-V(y-Y)\right)\rho-\frac{D}{\hbar^2}(X-Y)^2\rho.
\label{master2}
\end{align}
Since we wish to treat the incoming light particle as an approximate plane wave state and the target particle as stationary, it makes sense to work in momentum space for the reflected particle and position space for the target. The perturbative solution to (\ref{master2}) is given to second order in $V$ as
\begin{align}
\rho_t(p,q;X,Y) =  & \frac{1}{2\pi\hbar}\int_0^t dt' \int_0^{t'} dt'' \int  {dp_0}{dq_0}{dp_1}{dq_1}
\int dX_0 dY_0dX_1 dY_1dX_2 dY_2\nonumber\\
& \times \exp\left(-\frac{i(t-t')}{2m\hbar}(p^2-q^2)\right)J(X,Y,t|X_2,Y_2,t') \nonumber\\
& \times \frac{i}{\hbar}\left(V_{X_2}(p-p_1)\delta(q-q_1) - V^*_{Y_2}(q-q_1)\delta(p-p_1)\right)\nonumber\\
& \times \exp\left(-\frac{i(t'-t'')}{2m\hbar}(p_1^2-q_1^2)\right)J(X_2,Y_2,t'|X_1,Y_1,t'') \nonumber\\
& \times \frac{i}{\hbar}\left(V_{X_1}(p_1-p_0)\delta(q_1-q_0) - V^*_{Y_1}(q_1-q_0)\delta(p_1-p_0)\right)\nonumber\\
& \times \exp\left(-\frac{it''}{2m\hbar}(p_0^2-q_0^2)\right)J(X_1,Y_1,t''|X_0,Y_0,0) \rho_0(p_0,q_0; X_0, Y_0),
\label{pertMm}
\end{align}
where
\begin{align}
V_X(p) = \exp\left(-\frac{i}{\hbar}pX\right)V(p).
\end{align}
Here, we have ignored the zeroth and first order terms since they make no contribution to reflection.
We assume that the initial density matrix of the joint system is a plane wave for the light particle
and a steady state Gaussian for the target,
\begin{align}
\rho_0(p_0,q_0;   X_0, Y_0)\simeq \;& \frac{\sqrt{2\pi}\hbar}{\sigma}\delta(p_0 - \bar{p})\delta(q_0 - \bar{p})\nonumber\\
&\times \frac{1}{\sqrt[4]{2\pi\Sigma^2}}\exp\left(-\frac{(1-i)}{4\Sigma^2}(X_0-\bar{X})^2 + \frac{i}{\hbar}\bar{P}X_0\right)
\nonumber\\
&\times \frac{1}{\sqrt[4]{2\pi\Sigma^2}}\exp\left( -\frac{(1+i)}{4\Sigma^2}(Y_0-\bar{X})^2 - \frac{i}{\hbar}\bar{P}Y_0\right).
\label{initT}
\end{align}
We now evaluate Eq.(\ref{pertMm}) in a number of different cases.

\subsection{No environment}

We consider first the situation in which there is no environment present to see if the fluctuations in the target particle coming from its initial state are sufficient to suppress reflection of the light particle. This can be calculated either
from (\ref{pertMm}) with $D=0$, or directly from a simple perturbation theory
calculation in the unitary case. We find that
the probability of measuring the reflected particle with momentum $p$ and the target particle with momentum $P$ is
\begin{align}
        P_\infty^{\rm ref}(p,P) = & \rho_{\tau}^{\rm ref}(p,p; P, P) \nonumber\\
                 = & \frac{2\sqrt{\pi}\Sigma m}{\hbar^2\bar{p}}V^2(p-\bar{p})
                \exp\left(-\frac{\Sigma^2}{\hbar^2}(p-\bar{p}+P-\bar{P})^2\right)\nonumber\\
                 &\times \delta\left(\frac{\bar{p}^2}{2m}-\frac{p^2}{2m}+\frac{(P+p-\bar{p})^2}{2M}-\frac{P^2}{2M}\right),
\label{5.5}
\end{align}
(where we have made the usual assumption that the time integrals in Eq.(\ref{pertMm})
can be extended to infinity).
We can find the marginal probability of the reflected momentum $p$ of the light particle only by integrating over the $P$ variable. We find
\begin{align}
        P_\infty^{\rm ref}(p)  = & \frac{2\sqrt{\pi}\Sigma m M}{\hbar^2\bar{p}|p-\bar{p}|}V^2(p-\bar{p})
                \nonumber\\ &\times\exp\left(-\frac{\Sigma^2M^2}{\hbar^2 (p-\bar{p})^2}
\left[\frac{\bar{p}^2}{2m}-\frac{p^2}{2m}+\frac{\bar{P}^2}{2M}-\frac{(\bar{P}-p+\bar{p})^2}{2M}\right]^2\right).
\end{align}
The interpretation of this result is most easily seen by taking $ \bar P = 0$ and assuming
$m \ll M$, in which case we find,
\begin{align}
        P_\infty^{\rm ref}(p)  =  \frac{2\sqrt{\pi}\Sigma m M}{\hbar^2\bar{p}|p-\bar{p}|}V^2(p-\bar{p})
                 \exp\left(-\frac{\Sigma^2M^2}{4 \hbar^2 m^2} (p + \bar p)^2\right).
\end{align}
This shows the expected peak around $p = - \bar p $, but the peak will be very spread out
and flattened if
\beq
\frac {\bar p} {m}  \ll \frac{\Sigma_p}{M},
\eeq
where $\Sigma_p \sim \hbar / \Sigma $ is the momentum width of the initial target state. This is the condition
that the velocity fluctuations in the target are much larger than the incoming velocity, so this condition is of the same
form as those anticipated in the timescale analysis of Section 2.
This spreading will cause the total amount of reflected norm to be small, as discussed already.
Hence, an environment is not in fact necessary for an initial target state sufficiently spread out in momentum
and as long as the target final state is averaged out. Note that the large velocity fluctuations will require some sort of confining potential, as discussed in Section 2.

We now consider the more general case Eq.(\ref{5.5}) in which
the final target momentum is fixed. It is not hard to show that there are a wide variety of
final values of $P$ for which the reflection is still suppressed even without tracing it out. However, it is not true for {\it all} final target states and here we exhibit a situation in which reflection is not suppressed.

We can calculate a conditional probability density for measuring the reflected particle with momentum $p$ given that the target particle was measured to have momentum $P$.
This is given by
\begin{align}
        P_\infty^{\rm ref}(p|P) = \frac{P_\infty^{\rm ref}(p,P)}{P_\infty(P)}.
\end{align}
The probability for measuring the target particle with momentum $P$ is given to lowest order (no interaction) in the perturbative expansion as
\begin{align}
P_\infty(P) =& \int dP_0dQ_0J(P,P,\tau|P_0,Q_0,0)\rho_0(P_0,Q_0)\nonumber\\
                =&\frac{\Sigma}{\sqrt{\pi}}\frac{1}{\hbar}\exp\left(-\frac{\Sigma^2}{\hbar^2}(P-\bar{P})^2\right),
\end{align}
and so the conditional probability is given by
\begin{align}
        P_\infty^{\rm ref}(p|P) = & \frac{2\pi m}{\hbar\bar{p}}V^2(p-\bar{p})
                \exp\left(-\frac{\Sigma^2}{\hbar^2}\left[(p-\bar{p}+P-\bar{P})^2-(P-\bar{P})^2\right]\right)\nonumber\\
                 &\times \delta\left(\frac{\bar{p}^2}{2m}-\frac{p^2}{2m}+\frac{(P+p-\bar{p})^2}{2M}-\frac{P^2}{2M}\right).
        \label{condprob}
\end{align}
For example, it is perfectly reasonable to start with $\bar{P}=0$ and measure the final target momentum to be close to $P=0$. There will then be a $\delta$-function
peak around $p \approx - \bar p$ and for sufficiently small $\Sigma$ the exponential factor in (\ref{condprob}) will be $\sim 1$, so the probability will be
just like the unitary case for the usual scattering off a potential. There is no
spreading of the distribution so no suppression of reflection in this case.

\subsection{With environment}

We now consider the effects of the environment.
We first consider the situation in which the final state of the target is averaged over.
We have, of course, just shown that this case does not require an environment if the initial
target state is sufficiently broad in momentum, but the inclusion of the environment gives the
possibility of suppression without special choices of the initial state. There is also the
possibility that the environment will make the suppression of reflection more effective.

We consider the probability for reflection with momentum $p$, defined in terms of Eq.(\ref{pertMm}) by
\beq
P_\tau^{\rm ref}(p) = \int dX \rho_{\tau}^{\rm ref}(p,p; X, X).
\eeq
All the position and momentum integrals in (\ref{pertMm}) can be performed to leave
\begin{align}
P_\tau^{\rm ref}(p) =&  \frac{2m}{\hbar^2\bar{p}} V^2(p-\bar{p})
\frac{1}{\tau} \int_0^{\tau} ds \int_0^{\tau-s} du \nonumber\\
&\times \cos\left(\frac{s}{2m\hbar}(\bar{p}^2-p^2) + \frac{s}{2M\hbar}\left[\bar{P}^2-(\bar{P}-p+\bar{p})^2\right] \right) \nonumber\\
&\times \exp\left( - \frac{Ds^3}{3M^2\hbar^2}(p-\bar{p})^2- \frac{Ds^2u}{M^2\hbar^2}(p-\bar{p})^2- \frac{s^2}{4\Sigma^2M^2}(p-\bar{p})^2\right),
\label{resultpP}
\end{align}
where we only include combinations of potential terms which contribute to reflection. The $u$ integral can easily be proformed and the result is
\begin{align}
P_\tau^{\rm ref}(p) = & \frac{2 m}{\hbar^2\bar{p}} V^2(p-\bar{p}) \int_0^{\tau} ds \nonumber\\
&\times \cos\left(\frac{s}{2m\hbar}(\bar{p}^2-p^2) + \frac{s}{2M\hbar}\left[\bar{P}^2-(\bar{P}-p+\bar{p})^2\right] \right) \nonumber\\
&\times \frac{M^2\hbar^2}{Ds^2\tau (p-\bar{p})^2} \left[1 - \exp\left(- \frac{Ds^2(\tau-s)}{M^2\hbar^2}(p-\bar{p})^2\right)\right]\nonumber\\
&\times \exp\left( - \frac{Ds^3}{3M^2\hbar^2}(p-\bar{p})^2 - \frac{s^2}{4\Sigma^2M^2}(p-\bar{p})^2\right).
\label{5.12}
\end{align}
Note that in the limit $D\rightarrow 0$ we recover the
unitary result as expected.

For $D>0$ the last three exponential terms in Eq.(\ref{5.12}) indicate three distinct timescales which produce upper bound cutoffs to the $s$ integral. One can easily see that these three timescales are the Zeno time of the target $T_z$, the momentum decoherence
time of the target $T_d^p$, both defined in Section 2, and a new timescale,
\beq
T_1 = \frac {M\hbar } { \bar{p}\sqrt{Dt_z} } =  (T_d^p)^{3/2} t_z^{-1/2},
\eeq
where the Zeno time of the light particle $t_z$ defines the duration of the whole process. (Note also that these three timescales can also be read off from Eq.(\ref{resultpP})).
For reflection to be suppressed, we require that at least one of these timescales is much less than
the timescale $t_E$ in the cosine term, in order to spread out the resulting distribution, as described.
The requirement $T_z \ll t_E $ implies
\beq
T_{loc} \ll  \frac{m}{M} t_E,
\eeq
as discussed in Section 2, and this is equivalent to the relationship involving
velocities, Eq.(\ref{pres}).
The requirement $T_d^p \ll t_E $ implies
\beq
T_{loc} \ll \left( \frac {m}{M} \right)^{1/2} t_E.
\eeq
Note that this relation has a different power of $m/M$ compared to Eq.(\ref{tmom1})
and as a consequence is not a restriction on velocity fluctuations,
although this is not surprising, since this condition is about momentum decoherence which
is a different physical effect.

The requirement $T_1 \ll t_E $ implies
\beq
T_{loc} \ll \left( \frac{m}{M} t_E t_z\right)^{1/2}.
\eeq
This is the weakest of the three requirements, so it is the most important one, since
only one of them needs to be satisfied to suppress reflection. Note that again it has
a natural interpretation in terms of velocities:
\beq
\frac {p} {m} \ll \frac {(D t_z)^{1/2} } { M}.
\eeq
This means that the velocity of the light particle must be less than the velocity
fluctuations in the target that have accumluated by time $t_z$ due to evolution in
the presence of the environment.

We thus see that velocity fluctuations in the target are key for suppressing reflection.
Although here, unlike the unitary case, these fluctuations have been acquired though environmental interactions and not from the initial state.  Also note that momentum decoherence plays some role
in the suppression of reflection but the dominant effect appears to be the velocity
fluctuations in the target, and this is why
reflection can also be suppressed in the unitary case, with a suitable initial state for the target.

\begin{figure}[h]
        \begin{center}
                \includegraphics[width=13cm]{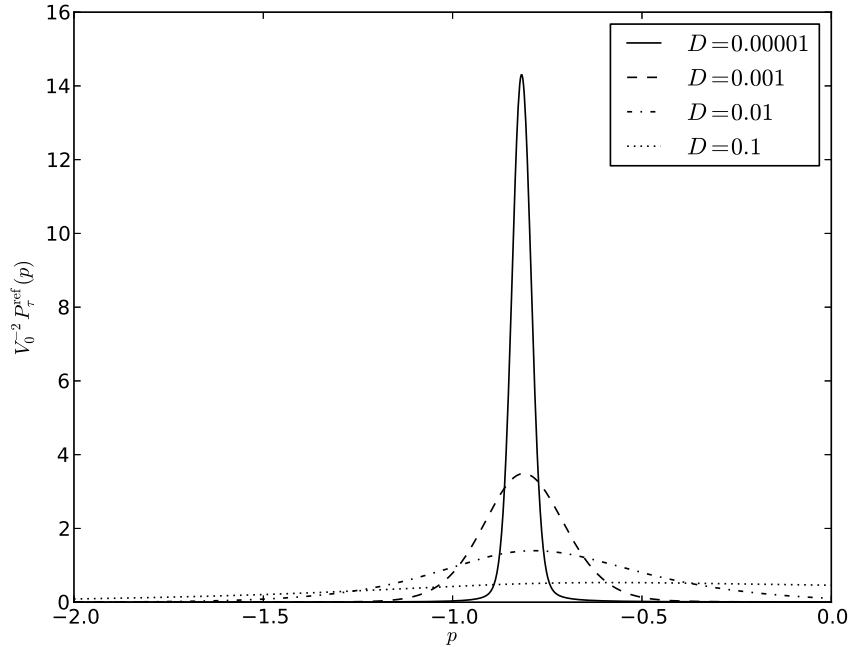}
        \caption{Plot showing the reflected probability density $P_{\tau}^{\rm ref}(p)$ as a function of momentum $p$ (see Eq.(\ref{5.12})) for Model II. The potential barrier is a Gaussian with associated length scale $a$. Plots are shown for a range of values of the diffusion parameter $D$. There is clear spreading and flattening of the peak for large values of $D$. Units are chosen such that $m = \bar{p} = \hbar = 1$ and we choose $a=0.1$, $\sigma = 100$ and $M=10$.}
\end{center}
\end{figure}

\begin{figure}[h]
        \begin{center}
                \includegraphics[width=13cm]{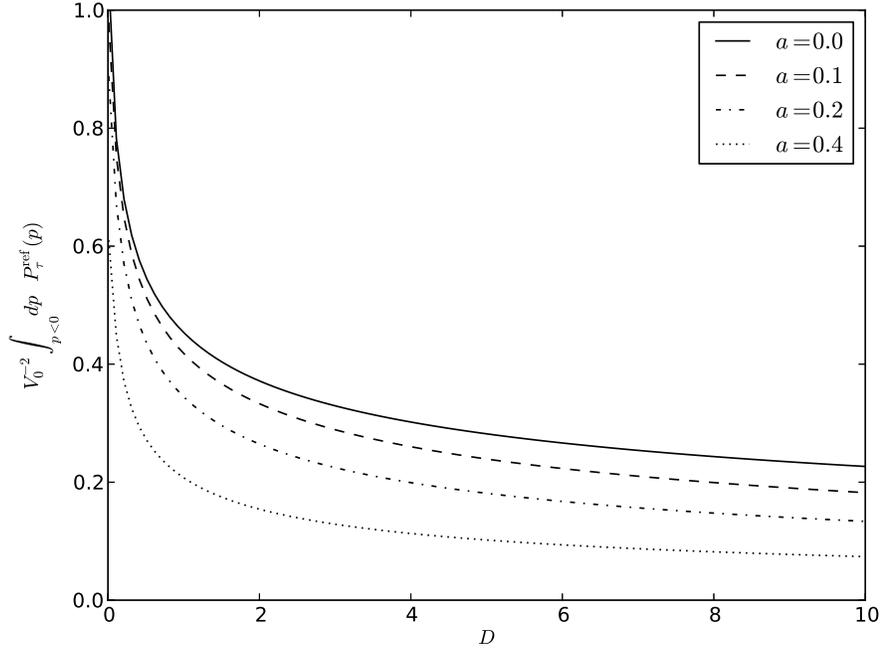}
        \caption{Plot showing the total reflected probability as a function of $D$ for Model II. The potential barrier is a Gaussian with associated length scale $a$. In the units chosen $D\sim 1$ sets a scale at which reflection in suppressed. Plots are shown for increasing values of $a$. Larger values of $a$ have the effect of decreasing the probability of reflection generally. Units are chosen such that $m = \bar{p} = \hbar = 1$ and we choose $\sigma = 100$ and $M=10$}.
\end{center}
\end{figure}

The reflected norm Eq.(\ref{5.12}) is plotted in Fig.(4). We see a clear spreading and flattening of the peak for large values of $D$. The total reflected norm is plotted in Fig.(5) for various values of $a$ and indicates suppression for large $D$ as expected. Note the close similarity with the $L \propto \hat p $ case for Model I plotted in Figs.(2), (3).

We finally consider the most general case in which the final state of the target is fixed but there is also an environment, which is necessary, since there is otherwise still reflection in this case for certain final target states.
The result for the conditional probability is
\begin{align}
        P_\tau^{\rm ref}(p|P) =&\frac{m}{\hbar^2\bar{p}}V^2(p-\bar{p})
                 \exp\left(-\frac{\Sigma^2}{4D\tau\Sigma^2+\hbar^2}\left[(p-\bar{p}+P-\bar{P})^2-(P-\bar{P})^2\right]\right)\nonumber\\
                & \times\frac{1}{\tau}\int_0^{\tau}ds\int_0^{\tau-s}du
                \exp\left(-\frac{is}{2m\hbar}\left(\bar{p}^2-p^2\right)\right)\nonumber\\
                &\times\exp\left(-\frac{is}{2M\hbar}\left[\bar{P}^2-(\bar{P}-p+\bar{p})^2\right]\right) \nonumber\\
                &\times\exp\left( -\frac{is}{2M\hbar}\frac{4D(s+2u)\Sigma^2+2\hbar^2}{4D\tau\Sigma^2+\hbar^2}
                (p-\bar{p})(p-\bar{p}+P-\bar{P})\right)\nonumber\\
                &\times \exp\left(- \frac{Ds^3}{3M^2\hbar^2}(p-\bar{p})^2- \frac{Ds^2u}{M^2\hbar^2}(p-\bar{p})^2\right)\nonumber\\
                &\times \exp\left(\frac{Ds^2}{4M^2\hbar^2}
                 \frac{4D(s+2u)^2\Sigma^2+4\hbar^2(s+2u-\tau)}{4D\tau\Sigma^2+\hbar^2}(p-\bar{p})^2\right)\nonumber\\
                & + \text{ complex conjugate}.
\end{align}
This is a complicated expression but its key features are easily seen and the analysis
is similar to the previous case. It is reasonable to take $ \tau \rightarrow \infty
$ in which case the final exponential tends to $\exp(-s^2(p-\bar{p})^2/4\Sigma^2M^2)$
and we then see that the last three exponential terms are identical to the last three
terms in Eq.(\ref{resultpP}), the key ones for our argument. We again therefore identify
the same three timescales and deduce that the reflected probability for any fixed final $P$ will be spread out, and thus reflection suppressed, under the same conditions.

\section{Summary and Discussion}

We have studied the general question of how quantum-mechanical reflection is suppressed in the quasi-classical limit using the standard machinery of decoherence through interaction with an environment. We first noted that the naive classical limit $\hbar \rightarrow 0 $ is not
sufficient although does highlight the role played by the smearing out of a potential with sharp edges. We then addressed the question of suppressing reflection in two different models.

In the first model, we considered an incoming light particle of positive momentum scattering off a barrier and classicalized the particle by coupling it to a thermal environment in a very standard way.
We found that the environment spreads out the reflected momentum, some of it into the positive momentum regime, some of it into the very large negative momentum regime where it is suppressed
mainly by the term coming from the smearing of the potential but also by prefactors in the scattering amplitude. The total amount of reflected momentum is therefore very small for a sufficiently strong environment. However, the associated fluctuations generated are so large that the incoming wave packet would acquire negative momenta at least as large as the reflection effect we are trying to suppress. We argued that this is due to the fact that the usual master equation with $L \propto \hat x$ decoheres very rapidly in position space but much more slowly in momentum space so is less effective in suppressing momentum interference effects, and quantum-mechanical reflection appears to be such an effect. This view was confirmed by using a different master equation with $L \propto \hat p $. Suppression of reflection was easily achieved in this case without large fluctuations but such a master equation is unphysical. We also analyzed the same system using the QSD approach which gave the same conclusions but from a rather different angle.

We thus arrive at the conclusion that quantum-mechanical reflection is not in fact suppressed by the obvious decoherence mechanism. This is somewhat surprising since this mechanism has been seen
in many situations to be highly effective in suppressing non-classical effects.

This conclusion also has implications for similar attempts to show that the quantum Zeno effect is suppressed by decoherence. We noted that exposing an incoming wave packet to a sequence of closely spaced projection operators onto the negative $x$-axis is approximately equivalent to evolution in a complex potential $ i V_0 \theta ( \hat x )$. The issue is then to argue that decoherence will suppress reflection from this potential, leaving only absorption (the classical limit of this process).
The arguments given in this paper, in particular the perturbative scattering calculation which is essentially identical, show that reflection cannot be suppressed without incurring unacceptably large fluctuations. This indicates problems in obtaining the classical limit of the Zeno effect in this case. This situation, which has further subtleties not discussed here, will be discussed in another paper \cite{BeHa2}.


The Zeno example also gives some clues as to the surprising ineffectiveness of environmental decoherence in this case. Reflection occurs when $V_0$ is of order $E$, the energy of the incoming particle, which means the time spacing $\e$ between projections is of order $ \hbar / E$. The point is that this is extremely close spacing of the projections, a regime that is generally not explored in most decoherence studies, in particular, in the decoherent histories approach where histories are typically characterized by projections that are quite widely spaced in time. That is, the situations in which decoherence works effectively are far from the Zeno regime considered here.
(Some related issues concerning decoherence and the Zeno effect are discussed in Ref.\cite{HaYePit}).

It was pointed out to us \cite{Bo} that
this ineffectiveness of decoherence in this case also has an interesting consequence.
Quantum-mechanical properties like interference and entanglement are usually suppressed extremely
effectively through decoherence and much effort is required experimentally to shield systems
from decoherence in order to preserve their quantum properties. Here we find that reflection is not easily suppressed. Reflection might therefore be regarded as a measure of ``quantumness'' which is relatively immune to decoherence. This will be explored further elsewhere.

Given the above difficulties, we turned to classicalization of the barrier as the possible source
of the suppression of reflection. We thus considered as a second model
a quantum-mechanical model of a barrier consisting of a massive target particle locally coupled to the light particle through a potential $V$ and chosen so that it reduced to the usual scattering problem when the target particle position and momentum were set to zero. To produce classicalization we added an environment which, crucially, was coupled to the target particle only, not the incoming light particle, so that large fluctuations of the incoming particle are avoided.
We found that quantum-mechanical reflection is suppressed under suitable conditions, essentially that the timescale of decoherence of the target particle (typically its localization time) is much less than the timescale of the scattering process (a combination of the energy time $t_E$ and Zeno time
$t_z$ of the light particle).  As in the first model, the reflection is suppressed because the reflected momentum is spread out into large positive and negative values where it is suppressed.

We also found that the environment is not always necessary for suppression of reflection in this way and in the case of no environment we found two results.
First, if the target particle is traced out in the scattering probability, the reflection probability of the light particle only is suppressed if its momentum is much less than the momentum fluctuations of the target. This is easily achieved if the target is put in an initial state which is very wide in momentum. Second, if the target particle is not traced out, there is still suppression of reflection of the light particle for a wide variety of final momenta of the target (under the same conditions on the
momentum), but not for all final target momenta. Hence the environment is necessary to ensure suppression of reflection for all possible final target momenta, the most general case.

We thus find in this second model in which a quantum barrier is classicalized that reflection is suppressed relatively easily. The essence of the mechanism is large velocity fluctuations in the barrier which get transferred to the incoming particle.

It is interesting to ask what would happen in the situation when {\it both} the incoming particle and quantum barrier are exposed to environmental interaction, as this would perhaps be the most natural situation to arise experimentally. This situation would be Model II with the additional feature that the environment couples to the incoming particle also. The analysis of Models I and II above allows us to see what would happen in this case. We found that a very strong environmental interaction with the incoming particle is required to suppress reflection but that the quantum barrier is classicalized and reflection suppressed relatively easily. Hence, it seems that a weakly interacting environment would be sufficient to suppress reflection in this modified Model II -- this would classicalize the quantum barrier quite easily
and have little effect on the incoming particle, so the analysis is essentially the same as Model II.
(This conclusion would, however, depend on the particular strength and nature of the coupling of the environment to each system and this would require a more specific model to assess in detail.)

We also mention here another model we have considered which is similar in flavour, consisting of scattering off a time-dependent barrier. We find that, in a lowest order perturbative calculation
of the scattering amplitude, reflection is suppressed if the barrier is oscillating in time at the appropriate frequency \cite{BeYe}.

In summary, quantum-mechanical reflection can be significantly suppressed through the classicalizing action of
environmental decoherence. It is achieved by the classicalization (and in particular the velocity
fluctuations) of the target or barrier off which the incoming particle scatters and not through the classicalization of the incoming particle.

\section{Acknowledgements}

We are grateful to Sougato Bose, James Yearsley and Wojtek Zurek for useful conversations.

\bibliography{apssamp}

\begin{thebibliography}{10}

\bibitem{Har6} J.B.Hartle, in, {\it Proceedings of
the Cornelius Lanczos International Centenary Confererence},
edited by J.D.Brown, M.T.Chu, D.C.Ellison and R.J.Plemmons (SIAM,
Philadelphia, 1994). Also available as e-print gr-qc/9404017.

\bibitem{JoZ} E.Joos and H.D.Zeh, Z. Phys. B59, 223 (1985).

\bibitem{Hal8} J.J.Halliwell, Phys. Rev. Lett. 83, 2481 (1999); Phys. Rev. D68, 025018 (2003).

\bibitem{Hal00} J.J.Halliwell, Contemp. Phys. 46, 93 (2005).


\bibitem{Zur1} W.H.Zurek, Phys. Today 44, 36 (1991); Rev. Mod. Phys. 75, 715 (2003);
Nature Physics, 5, 181 (2009); in {\it Physical Origins of Time Asymmetry},
edited by  J.J.Halliwell, J.Perez-Mercader and W.Zurek (Cambridge
University Press, Cambridge, 1994).

\bibitem{Zur2} W.Zurek, Phys. Rev. D24, 1516 (1981);
Phys. Rev. D26, 1862 (1982);
in {\it Frontiers of Nonequilibrium Statistical
Physics}, edited by G.T.Moore and M.O.Scully (Plenum, 1986).


\bibitem{Zur3} W.G.Unruh and W.Zurek, Phys. Rev. D40, 1071 (1989);
W.Zurek, S.Habib and J.P.Paz, Phys. Rev. Lett. 70,1187 (1993);
J.P.Paz and W.H.Zurek, Phys. Rev. D48, 2728 (1993).

\bibitem{BrHa} T.Brun and J.B.Hartle, Phys. Rev. D60, 123503 (1999).

\bibitem{GH2} M.Gell-Mann and J.B.Hartle, Phys. Rev. D47,
 3345 (1993).

\bibitem{GH1} M.Gell-Mann and J.B.Hartle, in {\it Complexity, Entropy
and the Physics of Information, SFI Studies in the Sciences of Complexity},
Vol. VIII, W. Zurek (ed.) (Addison Wesley, Reading, 1990); and in
{\it Proceedings of the Third International Symposium on the Foundations of
Quantum Mechanics in the Light of New Technology}, S. Kobayashi, H. Ezawa,
Y. Murayama and S. Nomura (eds.) (Physical Society of Japan, Tokyo, 1990).

\bibitem{Gri} R.B.Griffiths, J. Stat. Phys. 36, 219 (1984);
Phys. Rev. Lett. 70, 2201 (1993); Phys. Rev. A54, 2759
(1996); A57, 1604 (1998); {\it Consistent Quantum Theory} (Cambridge University Press,
Cambridge, 2002).

\bibitem{Omn} R.Omn\`es, J. Stat. Phys. 53, 893 (1988);
53, 933 (1988); 53, 957 (1988); 57, 357 (1989);
62, 841 (1991); Ann. Phys. 201, 354 (1990);
Rev. Mod. Phys. 64, 339 (1992); {\it The Interpretation of
Quantum Mechanics} (Princeton University Press, Princeton, 1994).

\bibitem{Hal1} J.J.Halliwell, in {\it Fundamental Problems in Quantum
Theory},  edited by D.Greenberger and A.Zeilinger, Annals of the
New York Academy of Sciences, 775, 726 (1994).

\bibitem{DoH} H.F.Dowker and J.J.Halliwell, Phys. Rev. D46, 1580 (1992).

\bibitem{HaDo} P.J.Dodd and J.J.Halliwell, Phys. Rev. A69, 052105 (2004).


\bibitem{LaLi} L.D.Landau and E.M.Lifschytz, Quantum Mechanics: Non-Relativistic Theory,
Third Edition (Pergamon Press, Oxford, 1977), section 25.

\bibitem{Lin} G.Lindblad, Comm. Math. Phys. 48, 119 (1976).



\bibitem{Wig} E.P.Wigner, Phys.Rev. 40, 749 (1932)
For extensive reviews of the Wigner function and related
functions, see N.L.Balazs and B.K.Jennings, Phys. Rep.
104, 347 (1984), and M.Hillery, R.F.O'Connell, M.O.Scully and E.P.Wigner, Phys. Rep. 106, 121 (1984).

\bibitem{KiNo} Y.S.Kim and M.E.Noz, Phase Space Picture of Quantum
Mechanics: Group Theoretical Approach, Lecture Notes in Physics Series Vol 40 (World Scientific, Singapore, 1991).

\bibitem{DiKi} L.Di\'osi and C.Kiefer, J. Math. Phys. A35, 2675
(2002).


\bibitem{ChTe} C.A.Chatzidimitriou-Dreismann and I.C.Teitje,
J. Phys: Conf Series 237, 012010 (2010); C.A.Chatzidimitriou-Dreismann
and S.Stenholm, quant-ph/0702038 (2007).


\bibitem{Zeno} B.Misra and E.C.G.Sudarshan, J. Math. Phys. 18, 756 (1977);  A. Peres,
Am. J. Phys. 48, 931 (1980).


\bibitem{Sch2} P. Facchi, S. Pascazio, A. Scardicchio, L. S. Schulman,
Phys. Rev. A65, 012108 (2002).


\bibitem{BeHa2} D.J.Bedingham and J.J.Halliwell,
{\it Suppression of the Zeno effect by decoherence}, in preparation.



\bibitem{Ech} J.Echanobe, A. del Campo and J.G.Muga, Phys. Rev. A77, 032112
(2008).

\bibitem{HaYe3} J.J.Halliwell and J.M.Yearsley, J. Phys. A43, 445303 (2010)



\bibitem{complex} J.G.Muga, S.Brouard, D.Macias, Ann. Phys. (NY) 240, 351 (1995);
Ph.Blanchard and A.Jadczyk, Helv. Phys. Acta 69, 613 (1996);
J.P.Palao, J.G.Muga, S.Brouard, A.Jadczyk, Phys. Lett. A233, 227
(1997); A.Ruschhaupt, J. Phys. A35, 10429 (2002);
J.G.Muga, J.P.Palao, B.Navarro, I.L.Egusquiza, Phys. Rep. 395, 357 (2004);
J.J.Halliwell, Prog. Th. Phys. 102, 707 (1999); Phys. Rev. A77, 062103 (2008).


\bibitem{time} J.G.Muga, R.Sala Mayato and I.L.Egusquiza (eds),
{\it Time in Quantum Mechanics} (Springer, Berlin, 2002); J.G.Muga and
C.R.Leavens, Phys. Rep. 338, 353 (2000).

\bibitem{All} G.R.Allcock, Ann. Phys. 53, 253 (1969);
53, 286 (1969); 53, 311 (1969).

\bibitem{HaYe1} J.J.Halliwell and J.M.Yearsley, Phys. Rev. A79, 062101 (2009).

\bibitem{HaYe2} J.J.Halliwell and J.M.Yearsley,
 Phys. Lett. A374, 154 (2009).




\bibitem{GP1} N.Gisin and I.C. Percival, J. Phys. A25,
5677 (1992); Phys. Lett. A167, 315 (1992).

\bibitem{GP2} N.Gisin and I.C.Percival, J. Phys. A26,
2233 (1993).

\bibitem{GP3} N.Gisin and I.C.Percival, J. Phys.
A26, 2245 (1993).




\bibitem{Dio8} L.Di\'osi, Phys. Lett. A122, 221 (1987).

\bibitem{Dio2} L.Di\'osi, Phys. Lett. A129, 419 (1988).

\bibitem{Dio3} L.Di\'osi, Phys. Lett. A132, 233 (1988).






\bibitem{DGHP} L.Di\'osi, N.Gisin, J.Halliwell and I.C.Percival,
Phys. Rev. Lett. 74, 203 (1995).

\bibitem{HaZo1} J.J.Halliwell and A.Zoupas, Phys. Rev. D52 7294 (1995).

\bibitem{HaZo2} J.J.Halliwell and A.Zoupas, Phys. Rev. D55, 4697 (1997).


\bibitem{CaLe} A.O.Caldeira and A.J.Leggett, Physica
121A, 587 (1983).


\bibitem{AnHa} C.Anastopoulos and J.J.Halliwell, Phys. Rev. D51, 6870 (1995).

\bibitem{Yeapp} J.M.Yearsley, private communication.

\bibitem{HaYePit} J.J.Halliwell and J.M.Yearsley, Phys. Rev. A86, 024016 (2012).

\bibitem{Bo} S.Bose, private communication.

\bibitem{BeYe} D.J.Bedingham and J.M.Yearsley, in preparation.

\bibitem{Hal2D} J.J.Halliwell, J. Phys. A40, 3067, (2007).




































\end{thebibliography}

\end{document}